\documentclass[aps, showpacs, showkeys, nofootinbib, floatfix]{revtex4}
\usepackage{color}
\usepackage{amssymb}
\usepackage{amsmath}
\usepackage{graphicx}

\begin{document}

\title{Cosmological constraints on interacting dark energy with redshift-space distortion after Planck data}

\author{Weiqiang Yang}
\author{Lixin Xu\footnote{lxxu@dlut.edu.cn}}

\affiliation{Institute of Theoretical Physics, School of Physics and Optoelectronic Technology, Dalian University of Technology, Dalian, 116024, People's Republic of China}

\begin{abstract}
The interacting dark energy model could propose a effective way to avoid the coincidence problem. In this paper, dark energy is taken as a fluid with a constant equation of state parameter $w_x$. In a general gauge, we could obtain two sets of different perturbation equations when the momentum transfer potential is vanished in the rest frame of dark matter or dark energy. There are many kinds of interacting forms from the phenomenological considerations, here, we choose $Q=3H\xi_x\rho_x$ which owns the stable perturbations in most cases. Then, according to the Markov Chain Monte Carlo method, we constrain the model by currently available cosmic observations which include cosmic microwave background radiation, baryon acoustic oscillation, type Ia supernovae, and $f\sigma_8(z)$ data points from redshift-space distortion. Jointing the geometry tests with the large scale structure information, the results show a tighter constraint on the interacting model than the case without $f\sigma_8(z)$ data. We find the interaction rate in 3$\sigma$ regions: $\xi_x=0.00372_{-0.00372- 0.00372-0.00372}^{+0.000768+0.00655+0.0102}$. It means that the recently cosmic observations favor a small interaction rate between the dark sectors, at the same time, the measurement of redshift-space distortion could rule out a large interaction rate in the 1$\sigma$ region.
\end{abstract}

\maketitle

\section{Introduction}

In March 2013, the \textit{Planck} Collaboration and European Space Agency publicly released the new and precise measurements of the cosmic microwave background (CMB) radiation in a wide range of multiples ($l<2500$) \cite{ref:Planck2013,ref:Planck2013-CMB,ref:Planck2013-params,ref:Planck2013-download}. There is no doubt that this data will improve the accuracy of constraining the cosmological models. After \textit{Planck} data, the CMB data sets include two main parts: one is the low-l (up to a maximum multipole number of $l=49$) and high-l (from $l=50$ to $l=2500$) temperature power spectrum likelihood from \textit{Planck} \cite{ref:Planck2013-params}; the other is the low-l (up to $l=32$) polarization power spectrum likelihood from nine-year Wilkinson Microwave Anisotropy Probe (WMAP9) \cite{ref:WMAP9}. The observational constraints on the standard model from the CMB data show us that the Universe is composed by $68\%$ dark energy, $28\%$ dark matter, and $4\%$ baryons \cite{ref:Planck2013-params}.

The \textit{Planck} data are in good agreement with the $\Lambda$CDM model which is composed by the cosmological constant and cold dark matter (CDM), especially for the high multiples ($l>40$). However, the standard scenario itself is encountering the coincidence problem \cite{ref:Zlatev1999,ref:Chimento2003,ref:Huey2006}, which points out the fact that there is no reasonable explanation why the energy densities of vacuum energy and dark matter are of the same order today. In order to avoid this issue, one direct way to is to describe dark energy as a fluid and consider its equation of state (EoS) $w_x$ as a free parameter. This model is usually called as the $w$CDM model. Constraints on this extensional model from the CMB and baryon acoustic oscillation (BAO) data sets present that $w_x=-1.13^{+0.24}_{-0.25}$ with 95$\%$ confidence levels (C.L.) \cite{ref:Planck2013-params}.

An alternative powerful mechanism to alleviate the coincidence problem is to consider the interaction between dark matter and dark energy. First, the standard model of particle physics thinks the interaction within the dark sectors could be a natural choice, the uncoupled case would be an additional assumption on some model \cite{ref:Peebles2010}. It is worth looking forward to obtain the concrete form of interaction from the first principles. However, this idea is scarcely possible because the physical nature of dark matter and dark energy are still unknown. In most cases, one could assume the form of interaction from the phenomenological considerations. A satisfactory interacting model at least requires that the interacting form $Q$ should be expressed with respect to the energy densities of dark fluids and other covariant quantities, some possibilities of the interaction between the dark sectors have been widely discussed in Refs. \cite{
ref:Marulli2012,ref:Baldi2012,ref:Baldi2011,
ref:Baldi2010,ref:Beynon2012,
ref:Lee2012,ref:Cui2012,ref:Xia2013,ref:Xia2009,ref:Pourtsidou2013,ref:Tarrant2012,ref:Ziaeepour2012,
ref:Beyer2011,ref:Souza2010,ref:Vacca2009,ref:Manera2006,ref:Corasaniti2008,ref:Mota2008,ref:Amendola2006,
ref:Amendola2004,ref:Amendola2000,ref:Amendola2000-2,ref:Valentini2002,ref:Amendola1999,ref:Wetterich1995,
ref:Holden2000,ref:Das2006,ref:Hwang2001,ref:Hwang2002,
ref:Potter2011,ref:Aviles2011,
ref:Caldera-Cabral2009,ref:Caldera-Cabral2009-2,ref:Boehmer2010,ref:Boehmer2008,ref:Song2009,ref:Koyama2009,
ref:Majerotto2010,ref:Valiviita2010,ref:Valiviita2008,ref:Jackson2009,ref:Clemson2012,
ref:Bean2008,ref:Bean2008-2, 
ref:Chongchitnan2009,ref:Gavela2009,ref:Gavela2010,ref:Salvatelli2013,ref:Quartin2008,ref:Honorez2010,ref:Costa2013,ref:Bernardis2011,ref:He2011,ref:He2010,
ref:He2008,ref:Abdalla2009,ref:Sadjadi2010,ref:Olivares2008,ref:Olivares2006,
ref:Olivares2005,ref:Sun2013,ref:Sadjadi2006,ref:Sadeghi2013,ref:Zhang2013,ref:Koivisto2005,ref:Simpson2011,
ref:Bertolami2007,ref:Avelino2012,ref:Quercellini2008,ref:Mohammadi2012,ref:Sharif2012,ref:Fu2012,ref:Li2011,
ref:Barrow2006,ref:Zimdahl2001,
ref:Lip2011d,ref:Chen2011,ref:Chen2009,ref:Koshelev2009,ref:Zhang2012,ref:Cao2011,ref:Guo2007,
ref:Liyh2013,ref:Bolotin2013,ref:Chimento2013,ref:Chimento2012RDE,ref:Chimento2011RDE,ref:Tong2011}. Roughly, we divide these works into three main types. Interacting model (I) is $Q=\beta\rho_c\dot{\varphi}$ or $Q=\beta(\varphi)\rho_c\dot{\varphi}$ \cite{
ref:Marulli2012,ref:Baldi2012,ref:Baldi2011,
ref:Baldi2010,ref:Beynon2012,
ref:Lee2012,ref:Cui2012,ref:Xia2013,ref:Xia2009,ref:Pourtsidou2013,ref:Tarrant2012,ref:Ziaeepour2012,
ref:Beyer2011,ref:Souza2010,ref:Vacca2009,ref:Manera2006,ref:Corasaniti2008,ref:Mota2008,ref:Amendola2006,
ref:Amendola2004,ref:Amendola2000,ref:Amendola2000-2,ref:Valentini2002,ref:Amendola1999,ref:Wetterich1995,
ref:Holden2000,ref:Das2006,ref:Hwang2001,ref:Hwang2002} which might be motivated within the context of scalar-tensor theories. Although model (I) could have a significant physical motivation, but it meets with a challenge \cite{ref:Amendola2006}: the accelerated scaling attractor is not connected to a matter era where structure grows in the standard way. Far from this defect, some other interacting models have been suggested and discussed.
Interacting model (II) is $Q=\Gamma_c\rho_c$, $Q=\Gamma_x\rho_x$, or $Q=\Gamma_c\rho_c+\Gamma_x\rho_x$ \cite{ref:Potter2011,ref:Aviles2011,
ref:Caldera-Cabral2009,ref:Caldera-Cabral2009-2,ref:Boehmer2010,ref:Boehmer2008,ref:Song2009,ref:Koyama2009,
ref:Majerotto2010,ref:Valiviita2010,ref:Valiviita2008,ref:Jackson2009,ref:Clemson2012} which is not in the light of physical interaction between the dark sectors but is assumed for mathematical simplicity. $\Gamma_c$ or $\Gamma_x$ is a constant interaction rate which is determined by local interactions. Furthermore, if one considers interaction could be influenced by the expansion rate $H$ of the Universe, interacting model (III) could be designed as $Q=3H\xi_c\rho_c$, $Q=3H\xi_x\rho_x$, or $Q=3H(\xi_c\rho_c+\xi_x\rho_x)$ \cite{ref:Chongchitnan2009,ref:Gavela2009,ref:Gavela2010,ref:Salvatelli2013,ref:Quartin2008,ref:Honorez2010,ref:Costa2013,ref:Bernardis2011,ref:He2011,ref:He2010,
ref:He2008,ref:Abdalla2009,ref:Sadjadi2010,ref:Olivares2008,ref:Olivares2006,
ref:Olivares2005,ref:Sun2013,ref:Sadjadi2006,ref:Sadeghi2013,ref:Zhang2013,ref:Koivisto2005,ref:Simpson2011,
ref:Bertolami2007,ref:Avelino2012,ref:Quercellini2008,ref:Mohammadi2012,ref:Sharif2012,ref:Fu2012,ref:Li2011,
ref:Barrow2006,ref:Zimdahl2001}. This kind of model could produce an accelerated scaling attractor which might be connected to a standard matter era \cite{ref:Olivares2008}. Apart from the three main types of interacting models, some other generalized interacting forms have been studied in Refs.
\cite{ref:Lip2011d,ref:Chen2011,ref:Chen2009,ref:Koshelev2009,ref:Zhang2012,ref:Cao2011,ref:Guo2007,
ref:Liyh2013,ref:Bolotin2013,ref:Chimento2013,ref:Chimento2012RDE,ref:Chimento2011RDE,ref:Tong2011}.

Interacting dark energy could exert a non-gravitational 'drag' on dark matter, which will influences the evolution of matter density perturbations and the expansion history of the Universe. It means that some new features could be introduced into structure formation \cite{ref:Baldi2011,ref:Clemson2012,ref:Caldera-Cabral2009,ref:Song2009,ref:Koyama2009,ref:Honorez2010,ref:Koshelev2009}.
So, in the process of exploring the interaction, it is necessary to consider the effects on the cosmological constraints from the large scale structure information. Moreover, comparing with the geometry information (CMB, BAO, and type Ia supernovae (SN)), the large scale structure information is a powerful tool to break the possible degeneracy of cosmological models, because the dynamical growth history of different models could be distinct even if they might undergo similar background evolution behavior. Based on the redshift-space distortion (RSD), the currently observed $f_{obs}$ data could be closely associated with the evolution of matter density perturbations $\delta_m$ via the relation $f_m=d\ln\delta_m/d\ln a$, but it depends on the $\Lambda$CDM model. To keep away from this disadvantage, Song and Percival suggested to constrain the dark energy models by use of the model-independent $f\sigma_8(z)$ measurement \cite{ref:fsigma8-DE-Song2009}, in which $\sigma_8$ is the root-mean-square mass fluctuation in spheres with radius $8h^{-1}$ Mpc. Inspired by this paper, Xu combined the $f\sigma_8(z)$ data with the geometry measurements to constrain the holographic dark energy model in Ref. \cite{ref:fsigma8-HDE-Xu2013}. After \textit{Planck} data, Xu compared the deviation of growth index $\gamma_L$ (the growth function is parameterized as $f=\Omega_m^{\gamma_L}$) in the Einstein's gravity theory and modified gravity theory \cite{ref:fsigma8andPlanck-MG-Xu2013} and confronted Dvali-Gabadadze-Porrati braneworld gravity with the RSD measurement \cite{ref:Xu2013-DGP}. Besides, Yang and Xu explored the possible existence of warm dark matter from $f\sigma_8(z)$ test \cite{ref:Yang2013-1}, and Yang \textit{et al.} constrained a decomposed dark fluid with constant adiabatic sound speed by combining the RSD data with the geometry tests \cite{ref:Yang2013-2}. All the above constraints on the cosmological models from the RSD test \cite{ref:fsigma8-HDE-Xu2013,ref:fsigma8andPlanck-MG-Xu2013,ref:Xu2013-DGP,ref:Yang2013-1,ref:Yang2013-2} obtained tighter constraints on the model parameter space than the case without the $f\sigma_8(z)$ data. Up to now, the ten observed data points of $f\sigma_8(z)$ are shown in Table \ref{tab:fsigma8data}.

\begin{table}
\begin{center}
\begin{tabular}{ccc}
\hline\hline z & $f\sigma_8(z)$ & Survey and Refs \\ \hline
$0.067$ & $0.42\pm0.06$ & $6dFGRS~(2012)$ \cite{ref:fsigma85-Beutler2012}\\
$0.17$ & $0.51\pm0.06$ & $2dFGRS~(2004)$ \cite{ref:fsigma81-Percival2004}\\
$0.22$ & $0.42\pm0.07$ & $WiggleZ~(2011)$ \cite{ref:fsigma82-Blake2011}\\
$0.25$ & $0.39\pm0.05$ & $SDSS~LRG~(2011)$ \cite{ref:fsigma83-Samushia2012}\\
$0.37$ & $0.43\pm0.04$ & $SDSS~LRG~(2011)$ \cite{ref:fsigma83-Samushia2012}\\
$0.41$ & $0.45\pm0.04$ & $WiggleZ~(2011)$ \cite{ref:fsigma82-Blake2011}\\
$0.57$ & $0.43\pm0.03$ & $BOSS~CMASS~(2012)$ \cite{ref:fsigma84-Reid2012}\\
$0.60$ & $0.43\pm0.04$ & $WiggleZ~(2011)$ \cite{ref:fsigma82-Blake2011}\\
$0.78$ & $0.38\pm0.04$ & $WiggleZ~(2011)$ \cite{ref:fsigma82-Blake2011}\\
$0.80$ & $0.47\pm0.08$ & $VIPERS~(2013)$ \cite{ref:fsigma86-Torre2013}\\
\hline\hline
\end{tabular}
\caption{The data points of $f\sigma_8(z)$ measured from RSD with the survey references. The former nine data points at $z\in[0.067,0.78]$ were summarized in Ref. \cite{ref:fsigma8total-Samushia2013}. The data point at $z=0.8$ was released by the VIPERS in Ref. \cite{ref:fsigma86-Torre2013}. Then, a lower growth rate from RSD than expected from \textit{Planck} was also pointed out in Ref. \cite{ref:fsigma87-Macaulay2013}.}
\label{tab:fsigma8data}
\end{center}
\end{table}

The interaction rate should be determined by the cosmic observations. Since \textit{Planck} data have been released, several interacting dark energy models have been constrained by the recently cosmic observations \cite{ref:Salvatelli2013,ref:Costa2013,ref:Liyh2013,ref:Chimento2013}. In our cosmological constraints, the CMB data is from \textit{Planck} \cite{ref:Planck2013-params} and WMAP9 \cite{ref:WMAP9}. We use the measured ratio of $r_s/D_v$ as a 'standard ruler' to adopt the BAO data, the concrete values at three different redshifts are, $r_s/D_V(z=0.106)=0.336\pm0.015$ \cite{ref:BAO-1}, $r_s/D_V(z=0.35)=0.1126\pm0.0022$ \cite{ref:BAO-2}, and $r_s/D_V(z=0.57)=0.0732\pm0.0012$ \cite{ref:BAO-3}. For the SN data, we use the SNLS3 data which is composed by 472 SN calibrated by SiFTO and SALT2 \cite{ref:SNLS3-1,ref:SNLS3-2,ref:SNLS3-3}. The geometry measurements slightly favor the interaction between dark matter and dark energy, meanwhile, the growth rate of dark matter perturbations possibly rules out large interaction rate which was pointed out in Ref. \cite{ref:Clemson2012}. This would allow the use of the large scale structure information, which would significantly improve the constraints on the interacting models. So, in this paper, we will try to add the RSD measurement to constrain the interacting model. It is worthwhile to anticipate that the large scale structure measurement will give a tight constraint on the parameter space.

The outline of this paper is as follows. In Sec. II, in the background evolution, the interaction between the dark sectors could lead to the changes in the effective EoS of dark energy. Then, in a general gauge, via choosing the rest frame of dark matter or dark energy, two sets of different perturbation equations could be given by the vanishing momentum transfer potential. Furthermore, the stability of the perturbations determines the interacting form $Q=3H\xi_x\rho_x$ as our research emphasis, and the model parameter $\xi_x$ is also called as the interaction rate in this paper. In Sec. III, when the interaction rate was varied, we showed the cosmological implications on the CMB temperature power spectra and matter power spectra. Moreover, we presented the modified growth of structure and evolution curves of $f\sigma_8(z)$. Based on the Markov Chain Monte Carlo (MCMC) method, we performed the cosmological constraints on the IwCDM model (the wCDM model with interaction between the dark sectors). Section IV is the conclusion.

\section{The background equations and perturbation equations}

When the interaction $Q$ between the dark sectors is considered, one can write the evolution equations for the energy densities of dark matter and dark energy as,
\begin{eqnarray}
\rho'_c+3\mathcal{H}\rho_c=aQ_c=-aQ,
\label{eq:rhoc}
\end{eqnarray}
\begin{eqnarray}
\rho'_x+3\mathcal{H}(1+w_x)\rho_x=aQ_x=aQ,
\label{eq:rhox}
\end{eqnarray}
where the prime denotes the derivative with respect to conformal time $\tau$ and the subscript c and x, respectively, stand for dark matter and dark energy, $w_x=p_x/\rho_x$ and $\mathcal{H}=d\ln a/d\tau$. $Q$ represents the rate of energy density transfer, so $Q>0$ means that the direction of energy transfer is from dark matter to dark energy, $Q<0$ implies the opposite situation. Based on the above two equations, we could define the effective EoS of dark matter and dark energy,
\begin{eqnarray}
w_{c,eff}=\frac{aQ}{3\mathcal{H}\rho_c},~~~~w_{x,eff}=w_x-\frac{aQ}{3\mathcal{H}\rho_x},
\label{eq:wceff-wxeff}
\end{eqnarray}
when we consider the dark energy as quintessence case ($w\geq-1$) and $Q>0$, the effective EoS $w_{x,eff}$ could cross the phantom divide ($w=-1$), this interacting quintessence behaves like an uncoupled 'phantom' model, moreover, does not have any negative kinetic energies. At the same time, the possible existence of this case might be influenced by the instability of the perturbations.

In a general gauge, the perturbed Friedmann-Robertson-Walker (FRW) metric is \cite{ref:Majerotto2010,ref:Valiviita2008,ref:Clemson2012}
\begin{eqnarray}
ds^2=a^2(\tau)\{ -(1+2\phi)d\tau^2+2\partial_iBd\tau dx^i+[(1-2\psi)\delta_{ij}+2\partial_i\partial_jE]dx^idx^j \},
\label{eq:per-metric}
\end{eqnarray}
where $\phi$, B, $\psi$ and E are the gauge-dependent scalar perturbations quantities.

The four-velocity of A fluid is given by \cite{ref:Majerotto2010,ref:Valiviita2008,ref:Clemson2012}
\begin{eqnarray}
u^{\mu}_A=a^{-1}(1-\phi,\partial^iv_A),
\label{eq:Q-vector}
\end{eqnarray}
where $v_A$ is the peculiar velocity potential whose relation with the volume expansion is $\theta_A=-k^2(v_A+B)$ in Fourier space \cite{ref:Valiviita2008,ref:Ma1995}.

After considering the interaction between the fluids, one knows that the energy-momentum conservation equation of A fluid reads \cite{ref:Majerotto2010,ref:Valiviita2008,ref:Clemson2012}
\begin{eqnarray}
\nabla_{\nu}T^{\mu\nu}_A=Q^{\mu}_A,~~~~\sum\limits_{\rm{A}}{Q_A^\mu}=0,
\label{eq:balance-equation}
\end{eqnarray}
where $T^{\mu\nu}_A$ represents the A-fluid energy-momentum tensor. When $\tilde{Q}_A$ and $F^{\mu}_A$, respectively, represent the energy and momentum transfer rate, relative to the four-velocity $u^{\mu}$, one has \cite{ref:Majerotto2010,ref:Valiviita2008,ref:Clemson2012}
\begin{eqnarray}
Q^{\mu}_A=\tilde{Q}_Au^{\mu}+F^{\mu}_A,
\label{eq:Q-vector}
\end{eqnarray}
where $\tilde{Q}_A=Q_A+\delta Q_A$ and $F^{\mu}_A=a^{-1}(0,\partial^if_A)$, $Q_A$ is the background term of the general interaction, and $f_A$ is a momentum transfer potential. The perturbed energy-momentum transfer four-vector can be split as \cite{ref:Majerotto2010,ref:Valiviita2008,ref:Clemson2012}
\begin{eqnarray}
Q^A_0=-a[Q_A(1+\phi)+\delta Q_A],~~~~Q^A_i=a\partial_i[Q_A(v+B)+f_A],
\label{eq:Q-component}
\end{eqnarray}

The perturbed energy and momentum balance equations are \cite{ref:Majerotto2010,ref:Valiviita2008}
\begin{eqnarray}
\delta \rho'_A+3\mathcal{H}(\delta \rho_A+\delta p_A)-3(\rho_A+p_A)\psi'-k^2(\rho_A+p_A)(v_A+E')
=aQ_A\phi+a\delta Q_A,
\label{eq:delta-rhoA}
\end{eqnarray}
\begin{eqnarray}
\delta p_A+[(\rho_A+p_A)(v_A+B)]'+4\mathcal{H}(\rho_A+p_A)(v_A+B)+(\rho_A+p_A)\phi-\frac{2}{3}k^2p_A\pi_A
=aQ_A(v+B)+af_A,
\label{eq:delta-PA}
\end{eqnarray}

Defining the density contrast $\delta_A=\delta\rho_A/\rho_A$ and considering $\pi_A=0$, one has the general evolution equations for density perturbations (continuity) and velocity perturbations (Euler) equations for A fluid \cite{ref:Majerotto2010,ref:Valiviita2008,ref:Clemson2012}
\begin{eqnarray}
\delta'_A+3\mathcal{H}(c^2_{sA}-w_A)\delta_A
+9\mathcal{H}^2(1+w_A)(c^2_{sA}-c^2_{aA})\frac{\theta_A}{k^2}
+(1+w_A)\theta_A-3(1+w_A)\psi'+(1+w_A)k^2(B-E')
\nonumber \\
=\frac{a}{\rho_A}(-Q_A\delta_A+\delta Q_A)
+\frac{aQ_A}{\rho_A}\left[\phi+3\mathcal{H}(c^2_{sA}-c^2_{aA})\frac{\theta_A}{k^2}\right],
\label{eq:general-deltaA}
\end{eqnarray}
\begin{eqnarray}
\theta'_A+\mathcal{H}(1-3c^2_{sA})\theta_A-\frac{c^2_{sA}}{(1+w_A)}k^2\delta_A
-k^2\phi
=\frac{a}{(1+w_A)\rho_A}[(Q_A\theta-k^2f_A)-(1+c^2_{sA})Q_A\theta_A],
\label{eq:general-thetaA}
\end{eqnarray}
where $c^2_{aA}$ is the adiabatic sound speed whose definition is $c^2_{aA}=p'_A/\rho'_A=w_x+w'_x/(\rho'_A/\rho_A)$, and $c^2_{sA}$ is the A-fluid physical sound speed in the rest frame, its definition is $c^2_{sA}=(\delta p_A/\delta\rho_A)_{restframe}$ \cite{ref:Valiviita2008,ref:Kodama1984,ref:Hu1998,ref:Gordon2004}. In order to avoid the unphysical instability, $c^2_{sA}$ should be taken as a non-negative parameter \cite{ref:Valiviita2008}.

Next, we need to specialize the energy and momentum transfer rate between the dark sectors. In order to find the perturbation equations which apply to the interacting models (II) and (III), first, we specialize the momentum transfer potential as the simplest physical choice which is zero in the rest frame of either dark matter or dark energy \cite{ref:Valiviita2008,ref:Koyama2009}. This leads to two cases of simple interacting model which include energy transfer four-vector parallel to the four-velocity of dark matter or dark energy. In the light of Refs. \cite{ref:Clemson2012,ref:Koyama2009}, the momentum transfer potential $f_A$ is
\begin{eqnarray}
k^2f_A=Q_A(\theta-\theta_c),~~for~~Q^{\mu}_A~\parallel~u^{\mu}_c,
\label{eq:fA-uc}
\end{eqnarray}
\begin{eqnarray}
k^2f_A=Q_A(\theta-\theta_x),~~for~~Q^{\mu}_A~\parallel~u^{\mu}_x,
\label{eq:fA-ux}
\end{eqnarray}
furthermore, introducing a simple parameter of 'choosing the momentum transfer' $b$ \cite{ref:Jackson2009}
\[b = \left\{ \begin{array}{l}
 1,~~for~~Q^{\mu}_A~\parallel~u^{\mu}_c, \\
 0,~~for~~Q^{\mu}_A~\parallel~u^{\mu}_x, \\
 \end{array} \right.\]
in the rest frame of dark matter or dark energy, the momentum transfer potential $f_A$ could be unified as
\begin{eqnarray}
k^2f_A=Q_A[b(\theta-\theta_c)+(1-b)(\theta-\theta_x)]=Q_A[\theta-b\theta_c-(1-b)\theta_x],
\label{eq:fA-b}
\end{eqnarray}

Substituting the above relation into Eqs. (\ref{eq:general-deltaA}) and (\ref{eq:general-thetaA}), the continuity and Euler equations of A fluid could be reduced to
\begin{eqnarray}
\delta'_A+3\mathcal{H}(c^2_{sA}-w_A)\delta_A
+9\mathcal{H}^2(1+w_A)(c^2_{sA}-c^2_{aA})\frac{\theta_A}{k^2}
+(1+w_A)\theta_A-3(1+w_A)\psi'+(1+w_A)k^2(B-E')
\nonumber \\
=\frac{a}{\rho_A}(-Q_A\delta_A+\delta Q_A)
+\frac{aQ_A}{\rho_A}\left[\phi+3\mathcal{H}(c^2_{sA}-c^2_{aA})\frac{\theta_A}{k^2}\right],
\label{eq:general-deltaA0-b}
\end{eqnarray}
\begin{eqnarray}
\theta'_A+\mathcal{H}(1-3c^2_{sA})\theta_A-\frac{c^2_{sA}}{(1+w_A)}k^2\delta_A-k^2\phi
=\frac{aQ_A}{(1+w_A)\rho_A}[b\theta_c+(1-b)\theta_x-(1+c^2_{sA})\theta_A],
\label{eq:general-thetaA-b}
\end{eqnarray}

For the IwCDM model, $c^2_{sc}=c^2_{ac}=w_c=0=w'_x$ and $c^2_{ax}=w_x$, so the continuity and Euler equations become
\begin{eqnarray}
\delta'_x+3\mathcal{H}(c^2_{sx}-w_x)\delta_x
+9\mathcal{H}^2(1+w_x)(c^2_{sx}-w_x)\frac{\theta_x}{k^2}
+(1+w_x)\theta_x-3(1+w_x)\psi'+(1+w_x)k^2(B-E')
\nonumber \\
=\frac{a}{\rho_x}(-Q_x\delta_x+\delta Q_x)+
\frac{aQ_x}{\rho_x}\left[\phi+3\mathcal{H}(c^2_{sx}-w_x)\frac{\theta_x}{k^2}\right],
\label{eq:general-deltax-b}
\end{eqnarray}
\begin{eqnarray}
\delta'_c+\theta_c-3\psi'+k^2(B-E')
=-\frac{a}{\rho_c}\left(Q_c\delta_c-\delta Q_c\right)
+\frac{aQ_c}{\rho_c}\phi,
\label{eq:general-deltac-b}
\end{eqnarray}
\begin{eqnarray}
\theta'_x+\mathcal{H}(1-3c^2_{sx})\theta_x-\frac{c^2_{sx}}{(1+w_x)}k^2\delta_x-k^2\phi
=\frac{aQ_x}{(1+w_x)\rho_x}[b\theta_c+(1-b)\theta_x-(1+c^2_{sx})\theta_x],
\label{eq:general-thetax-b}
\end{eqnarray}
\begin{eqnarray}
\theta'_c+\mathcal{H}\theta_c-k^2\phi
=-\frac{aQ_c}{\rho_c}(1-b)(\theta_c-\theta_x),
\label{eq:general-thetac-b}
\end{eqnarray}

When the interaction is introduced, the instability of the perturbations becomes an important topic
\cite{ref:Valiviita2008,ref:Jackson2009,ref:Clemson2012,ref:Bean2008,ref:Bean2008-2,ref:Chongchitnan2009,
ref:Gavela2009}. In most cases, the energy transfer rate $Q=\Gamma_x\rho_x$ or $Q=3H\xi_x\rho_x$ owns the stable perturbations \cite{ref:Valiviita2008,ref:Clemson2012,ref:Gavela2009}. In this paper, we will choose the interacting model (III) as our research emphasis, so we take the interacting form as $Q=3H\xi_x\rho_x$. So, we have $Q_x=-Q_c=3H\xi_x\rho_x$ and $\delta Q_x=-\delta Q_c=3H\xi_x\rho_x\delta_x$. At the moment, the continuity and Euler equations could be recast into
\begin{eqnarray}
\delta'_x+3\mathcal{H}(c^2_{sx}-w_x)\delta_x
+9\mathcal{H}^2(1+w_x)(c^2_{sx}-w_x)\frac{\theta_x}{k^2}
+(1+w_x)\theta_x-3(1+w_x)\psi'+(1+w_x)k^2(B-E')
\nonumber \\
=3\mathcal{H}\xi_x\phi
+9\mathcal{H}^2(c^2_{sx}-w_x)\xi_x\frac{\theta_x}{k^2},
\label{eq:Clemson-deltax-b}
\end{eqnarray}
\begin{eqnarray}
\delta'_c+\theta_c-3\psi'+k^2(B-E')
=3\mathcal{H}\xi_x\frac{\rho_x}{\rho_c}(\delta_c-\delta_x)
-3\mathcal{H}\xi_x\frac{\rho_x}{\rho_c}\phi,
\label{eq:Clemson-deltac-b}
\end{eqnarray}
\begin{eqnarray}
\theta'_x+\mathcal{H}(1-3c^2_{sx})\theta_x-\frac{c^2_{sx}}{(1+w_x)}k^2\delta_x-k^2\phi
=\frac{3\mathcal{H}\xi_x}{1+w_x}
[b(\theta_c-\theta_x)-c^2_{sx}\theta_x],
\label{eq:Clemson-thetax-b}
\end{eqnarray}
\begin{eqnarray}
\theta'_c+\mathcal{H}\theta_x-k^2\phi
=3\mathcal{H}\xi_x\frac{\rho_x}{\rho_c}(1-b)(\theta_c-\theta_x),
\label{eq:Clemson-thetac-b}
\end{eqnarray}

Moreover, one could judge the stability of the perturbations via the doom factor \cite{ref:Gavela2009}. Here, we also define the doom factor for our IwCDM model
\begin{eqnarray}
d\equiv\frac{-Q}{3H\rho_x(1+w_x)}=\frac{-\xi_x}{1+w_x},
\label{eq:doom-factor}
\end{eqnarray}
according to the conclusion of Refs. \cite{ref:Gavela2009,ref:Clemson2012}: when $d<0$, the stable perturbations could be acquired for the interacting form $Q=3H\xi_x\rho_x$. It means that the perturbation stability requires the conditions $\xi_x>0$ and $(1+w_x)>0$ or $\xi_x<0$ and $(1+w_x)<0$. Here, in order to avoid the phantom doomsday \cite{ref:Caldwell2003}, we would discuss the stable case of $\xi_x>0$ and $(1+w_x)>0$.

In the
synchronous gauge ($\phi=B=0$, $\psi=\eta$, and $k^2E=-h/2-3\eta$), we rewrite the continuity and Euler equations as
\begin{eqnarray}
\delta'_x+(1+w_x)\left(\theta_x+\frac{h'}{2}\right)+3\mathcal{H}(c^2_{sx}-w_x)\delta_x
+9\mathcal{H}^2(1+w_x)(c^2_{sx}-w_x)\frac{\theta_x}{k^2}
=9\mathcal{H}^2(c^2_{sx}-w_x)\xi_x\frac{\theta_x}{k^2},
\label{eq:deltax-prime-b}
\end{eqnarray}
\begin{eqnarray}
\delta'_c+\theta_c+\frac{h'}{2}=3\mathcal{H}\xi_x\frac{\rho_x}{\rho_c}(\delta_c-\delta_x),
\label{eq:deltac-prime-b}
\end{eqnarray}
\begin{eqnarray}
\theta'_x+\mathcal{H}(1-3c^2_{sx})\theta_x-\frac{c^2_{sx}}{1+w_x}k^2\delta_x
=\frac{3\mathcal{H}\xi_x}{1+w_x}[b(\theta_c-\theta_x)-c^2_{sx}\theta_x],
\label{eq:thetax-prime-b}
\end{eqnarray}
\begin{eqnarray}
\theta'_c+\mathcal{H}\theta_c
=3\mathcal{H}\xi_x\frac{\rho_x}{\rho_c}(1-b)(\theta_c-\theta_x),
\label{eq:thetac-prime-b}
\end{eqnarray}



\section{Cosmological implications and constraint results}

\subsection{Theoretical predictions of CMB temperature and matter power spectra}

\begin{figure}[!htbp]
\includegraphics[width=13cm,height=9cm]{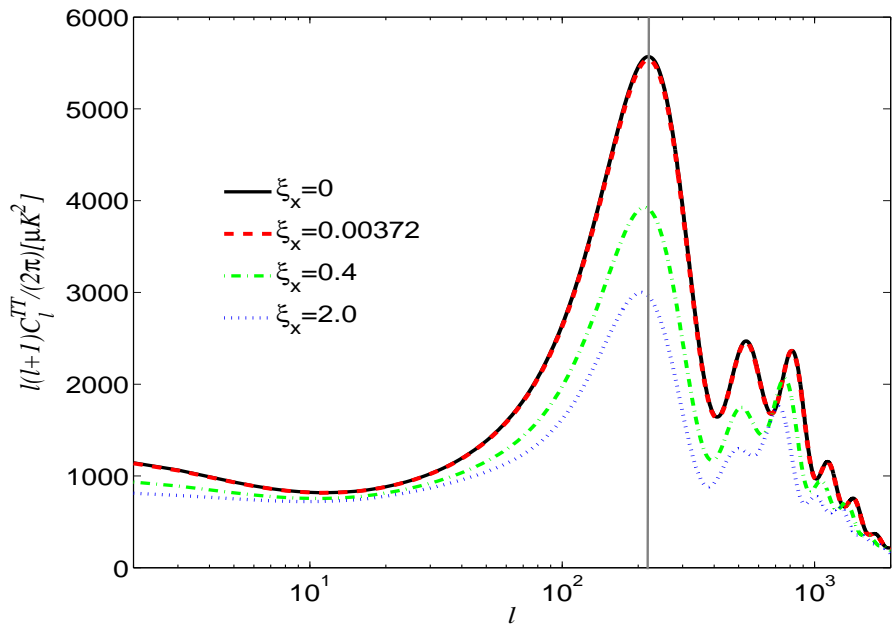}
  \caption{The effects on CMB temperature power spectra for the different values of interaction rate $\xi_x$. The black solid, red thick dashed, green dotted-dashed, and blue dotted lines are for $\xi_x=0, 0.00372, 0.4$, and $2.0$, respectively; the gray vertical line is used to clearly look into the shift tendency of the first peak; the other relevant parameters are fixed with the mean values as shown in the fifth column of Table \ref{tab:results-mean-pos}.}
  \label{fig:CMBpower-pos}
\end{figure}

\begin{figure}[!htbp]
\includegraphics[width=11cm,height=8cm]{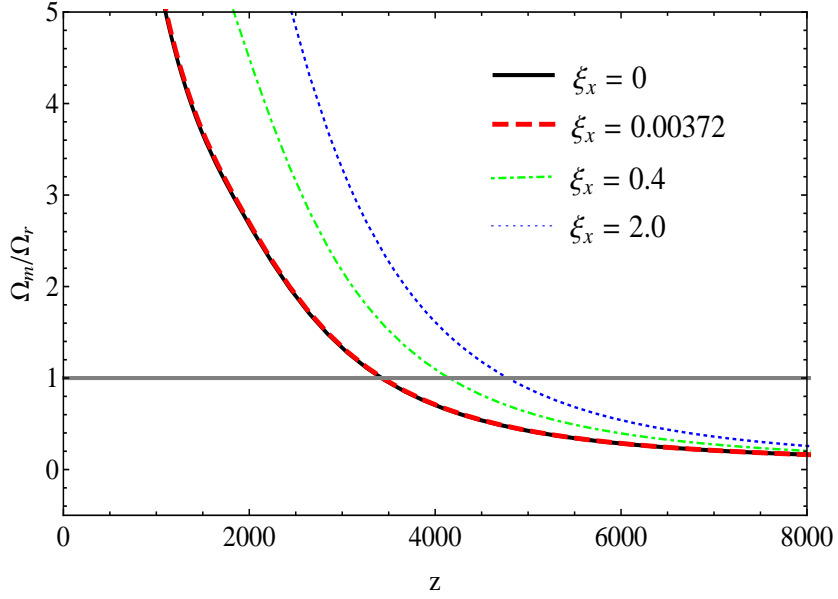}
  \caption{The evolutions for the ratio of dark fluid and radiation $\Omega_m/\Omega_r$ when the parameter $\xi_x$ is varied. The different lines correspond to the cases of the Fig. \ref{fig:CMBpower-pos}; the horizontal gray thick line responds to the case of $\Omega_m=\Omega_r$, and the other relevant parameters are fixed with the mean values as shown in the fourth column of Table \ref{tab:results-mean-pos}.}
  \label{fig:Omega-mr}
\end{figure}

\begin{figure}[!htbp]
\includegraphics[width=13cm,height=9cm]{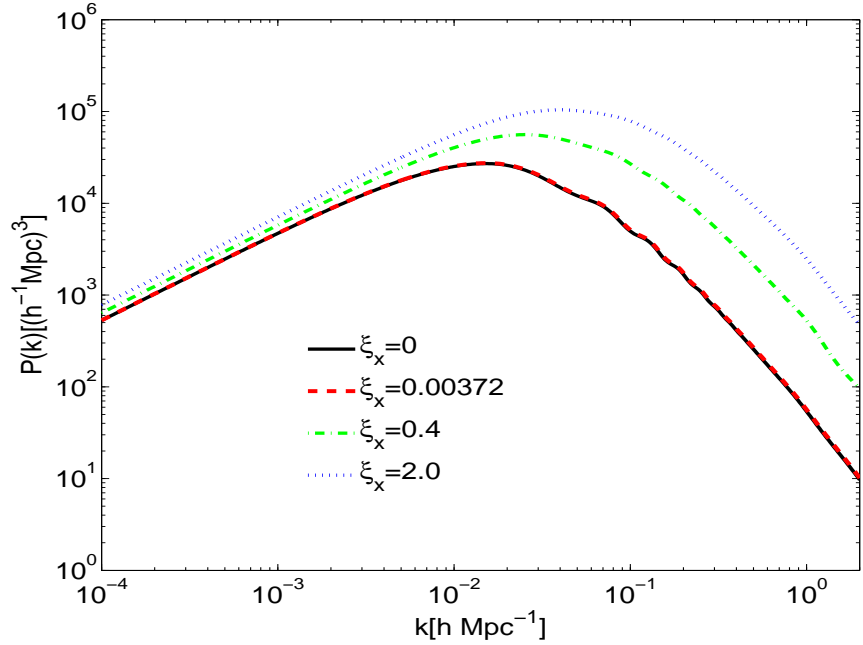}
  \caption{The effects on matter power spectra for the different values of interaction rate $\xi_x$. The black solid, red thick dashed, green dotted-dashed, and blue dotted lines are for $\xi_x=0, 0.00372, 0.4$, and $2.0$, respectively; the other relevant parameters are fixed with the mean values as shown in the fifth column of Table \ref{tab:results-mean-pos}.}
  \label{fig:Mpower-pos}
\end{figure}

When the interaction between the dark sectors is considered, some cosmological effects could take place, so we try to look for theoretical predictions of CMB temperature power spectra, matter power spectra, and the evolution curves of $f\sigma_8(z)$. Here, the cosmological implications have been discussed under the stability condition of the perturbations.
When the interaction rate $\xi_x$ is varied, the influences on the CMB temperature power spectra are presented in Fig. \ref{fig:CMBpower-pos}. In order to clearly show the relation between the interaction rate $\xi_x$ and the moment of matter-radiation equality, we also plot the evolution curves of $\Omega_m/\Omega_r$ in Fig. \ref{fig:Omega-mr}. From these two figures, we know that increasing the interaction rate $\xi_x$ is equivalent to enlarging the density parameter of effective dark matter $\Omega_m$, which could make the moment of matter-radiation equality earlier;
hence, the sound horizon is decreased. As a result, the first peak of CMB temperature power spectra is depressed.
As for the location shift of peaks, following the analysis about location of the CMB power spectra peaks on Ref. \cite{ref:Hu1995}, since the increasing $\xi_x$ is equivalent to enlarging $\Omega_m$, the peaks of power spectra would be shifted to smaller $l$. The similar case has occurred in Ref. \cite{ref:Clemson2012}. Moreover, since the shift of first peak is not significant, a vertical line could be used to clearly look into the shift tendency.
At large scales $l<100$, the integrated Sachs-Wolfe (ISW) effect is dominant, the changed parameter $\xi_x$ affects the CMB power spectra via ISW effect due to the evolution of gravitational potential. In Fig. \ref{fig:Mpower-pos}, we plot the influence on the matter power spectrum $P(k)$ for the different values of interaction rate $\xi_x$. The evolution law is opposite to the CMB temperature power spectra. With increasing the values of $\xi_x$, the matter power spectra $P(k)$ are enhanced due to the earlier matter-radiation equality. The case of $\xi_x=0.00372$ (corresponds the IwCDM model with mean value) and that of $\xi_x=0$ (corresponds to the uncoupled wCDM model) are almost the same.

\subsection{Modified growth of structure}

Based on the continuity and Euler equations of dark matter (\ref{eq:deltac-prime-b}), (\ref{eq:thetac-prime-b}), and $\ddot{h}+2H\dot{h}=-8\pi G(\delta\rho+3\delta p)$, we consider dark energy does not cluster on sub-Hubble scales \cite{ref:Koyama2009,ref:Clemson2012}, we could ignore the term $\delta_x$ in Eq. (\ref{eq:deltac-prime-b}) and obtain the second-order differential equation of density perturbation about dark matter

\begin{eqnarray}
\delta''_c+\left(1-3\xi_x\frac{\rho_x}{\rho_c}\right)\mathcal{H}\delta'_c
=4\pi Ga^2\rho_b\delta_b + 4\pi Ga^2\rho_c\delta_c \left\{1+ 2\xi_x\frac{\rho_{t}}{\rho_c}\frac{\rho_x}{\rho_c}
\left[ \frac{\mathcal{H}'}{\mathcal{H}^2}+1-3w_x+3\xi_x\left(1+\frac{\rho_x}{\rho_c}\right) \right] \right\}.
\label{eq:thetac-prime2-ide2-mod}
\end{eqnarray}
where $\mathcal{H}^2=8\pi Ga^2\rho_{t}/3$, $\rho_{t}=\rho_{r}+\rho_{b}+\rho_{c}+\rho_x$, the subscript $i=r,b,c,x$ respectively stand for radiation, baryons, dark matter and dark energy. When $\xi_x$=0, the above equation could be turned into the standard evolution of matter perturbations $\ddot{\delta}_m+\mathcal{H}\dot{\delta}_m=4\pi Ga^2\rho_m\delta_m$ \cite{ref:Linder2003}. This modification of the standard evolution for $\delta_c$ is different from the one of Ref. \cite{ref:Koyama2009} or Ref. \cite{ref:Clemson2012}, because the interacting form is different, particularly, in this paper, the energy exchange includes the expansion rate of the Universe.

The evolutions of $\delta_c$ for interacting model bring about the deviations from the standard evolutions of dark matter from two aspects. The first one is the modified effective expansion history $\mathcal{H}_{eff}$ in the background, that is, modified Hubble friction term; The second one is the modified effective gravitational constant $G_{eff}$, that is, modified source term, it might also be useful for distinguishing between IDE and modified gravity models \cite{ref:Clemson2012,ref:Tsujikawa2010}. Comparing with the standard equation of matter density perturbations, we could know
\begin{eqnarray}
\frac{\mathcal{H}_{eff}}{\mathcal{H}}=1-3\xi_x\frac{\rho_x}{\rho_c}.
\label{eq:Heff}
\end{eqnarray}

\begin{eqnarray}
\frac{G_{eff}}{G}=1+ 2\xi_x\frac{\rho_{t}}{\rho_c}\frac{\rho_x}{\rho_c}
\left[ \frac{\mathcal{H}'}{\mathcal{H}^2}+1-3w_x+3\xi_x\left(1+\frac{\rho_x}{\rho_c}\right) \right].
\label{eq:Geff}
\end{eqnarray}

\begin{figure}[!htbp]
\includegraphics[width=8.8cm,height=6.8cm]{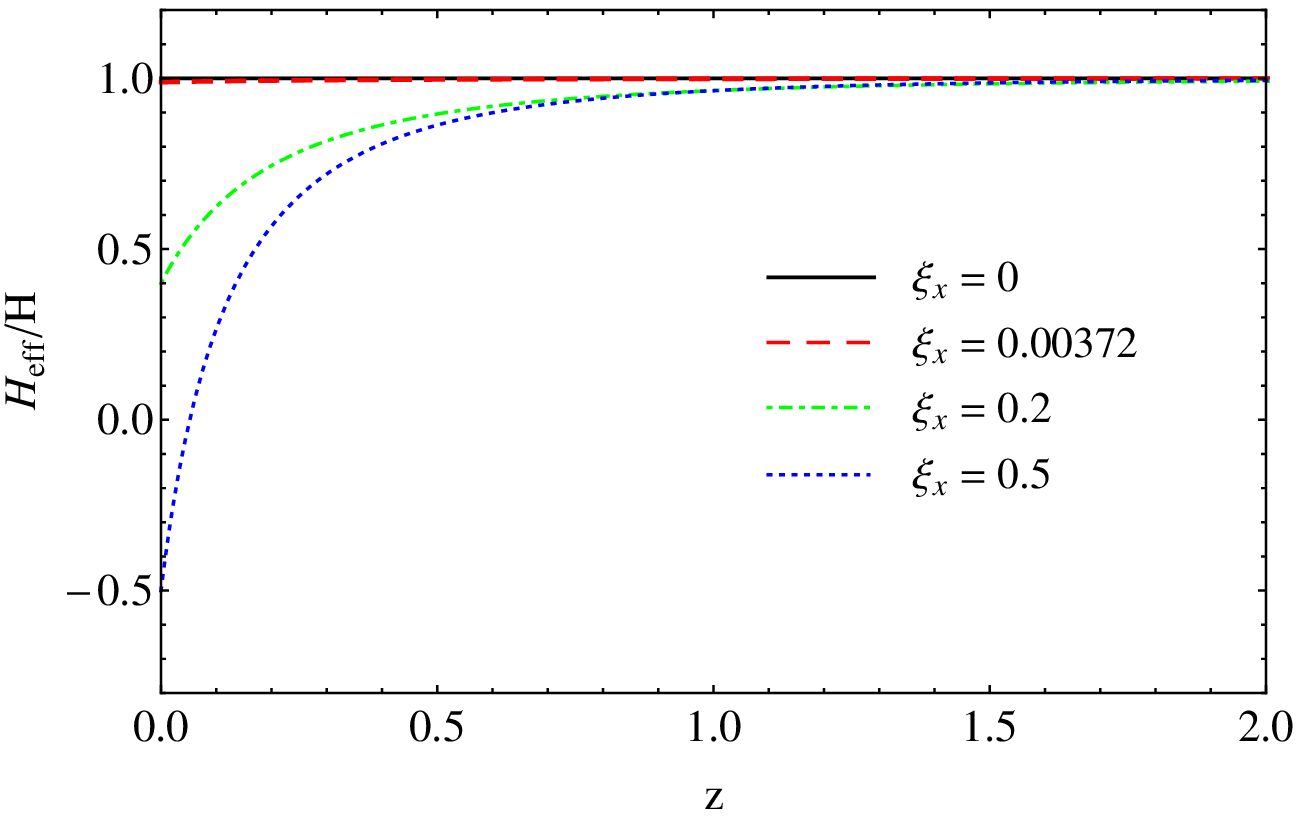}
\includegraphics[width=8.8cm,height=6.8cm]{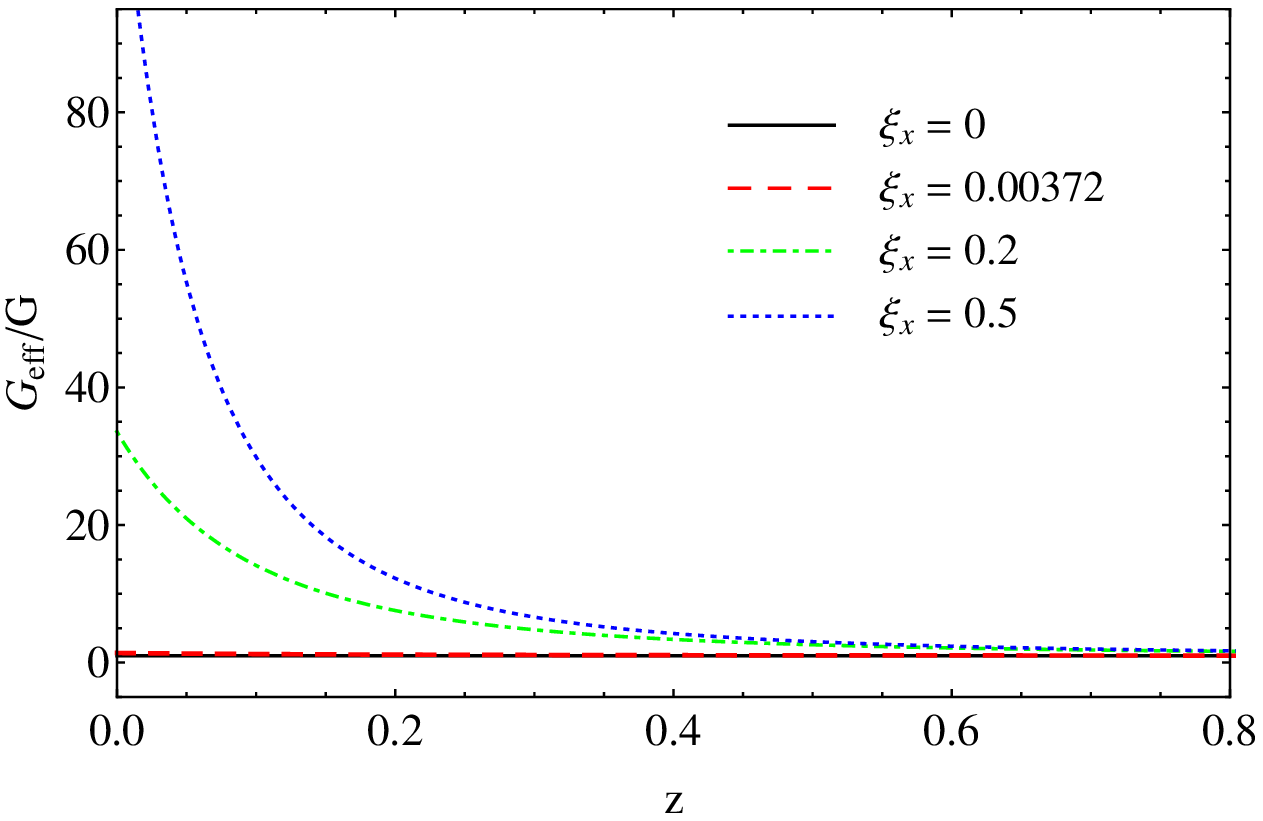}
  \caption{Deviations from standard model of the effective Hubble parameter (left panel) and effective Newton constant (right panel) for $\delta_c$. The black solid, red thick dashed, green dotted-dashed, and blue dotted lines are for $\xi_x=0, 0.00372, 0.4$, and $2.0$, respectively; $\xi_x=0$ corresponds to the case of $\Lambda$CDM model; the other relevant parameters are fixed with the mean values as shown in the fifth column of Table \ref{tab:results-mean-pos}.}
  \label{fig:Heff-Geff}
\end{figure}

The deviations from standard model of the effective Hubble parameter and effective gravitational constant for $\delta_c$ have been presented in Fig. \ref{fig:Heff-Geff}. With large interaction rate, from the past to today, the evolutions of $H_{eff}/H$ for $\delta_c$ take on exponential decreasement, meanwhile, the one of $G_{eff}/G$ shows exponential increasement.

\begin{figure}[!htbp]
\includegraphics[width=11cm,height=8cm]{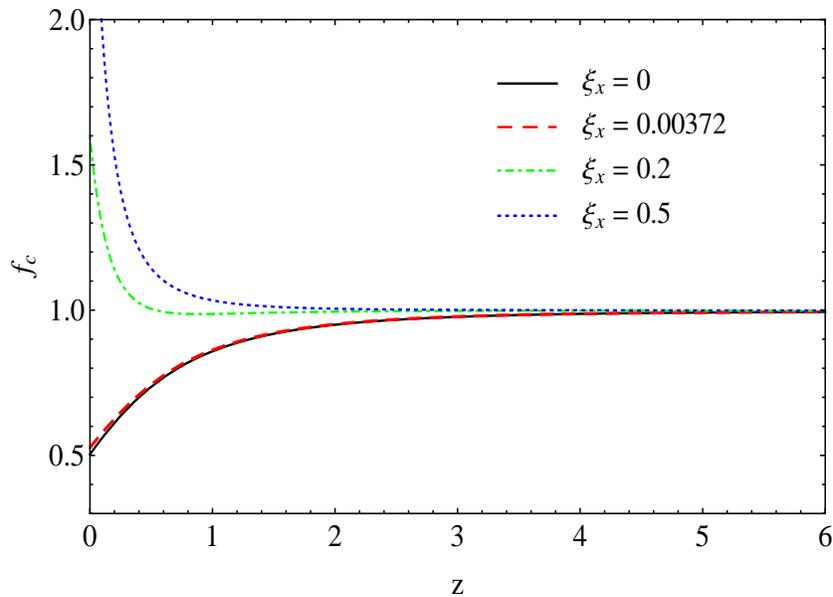}
  \caption{The evolutions for the growth rate of dark matter. The black solid, red thick dashed, green dotted-dashed, and blue dotted lines are for $\xi_x=0, 0.00372, 0.2$, and $0.5$, respectively; $\xi_x=0$ corresponds to the case of standard model; the other relevant parameters are fixed with the mean values as shown in the fifth column of Table \ref{tab:results-mean-pos}.}
  \label{fig:Dc-fc}
\end{figure}

As is known, the growth rate is $f_c=d\ln\delta_c/d\ln a$, modified evolution of $\delta_c$ determines that the growth history would deviate from the standard case in the theoretical frame of general relativity. When the interaction rate $\xi_x$ is varied, the evolutions of growth function and growth rate is shown in Fig. \ref{fig:Dc-fc}. From this figure, we clearly see that the interaction rate $\xi_x$ could significantly affect the growth history of the Universe, the growth rate presents large differences at late times. It means that the growth history of dark matter is significantly sensitive to the varied interaction rate.

\begin{figure}[!htbp]
\includegraphics[width=8.8cm,height=6.8cm]{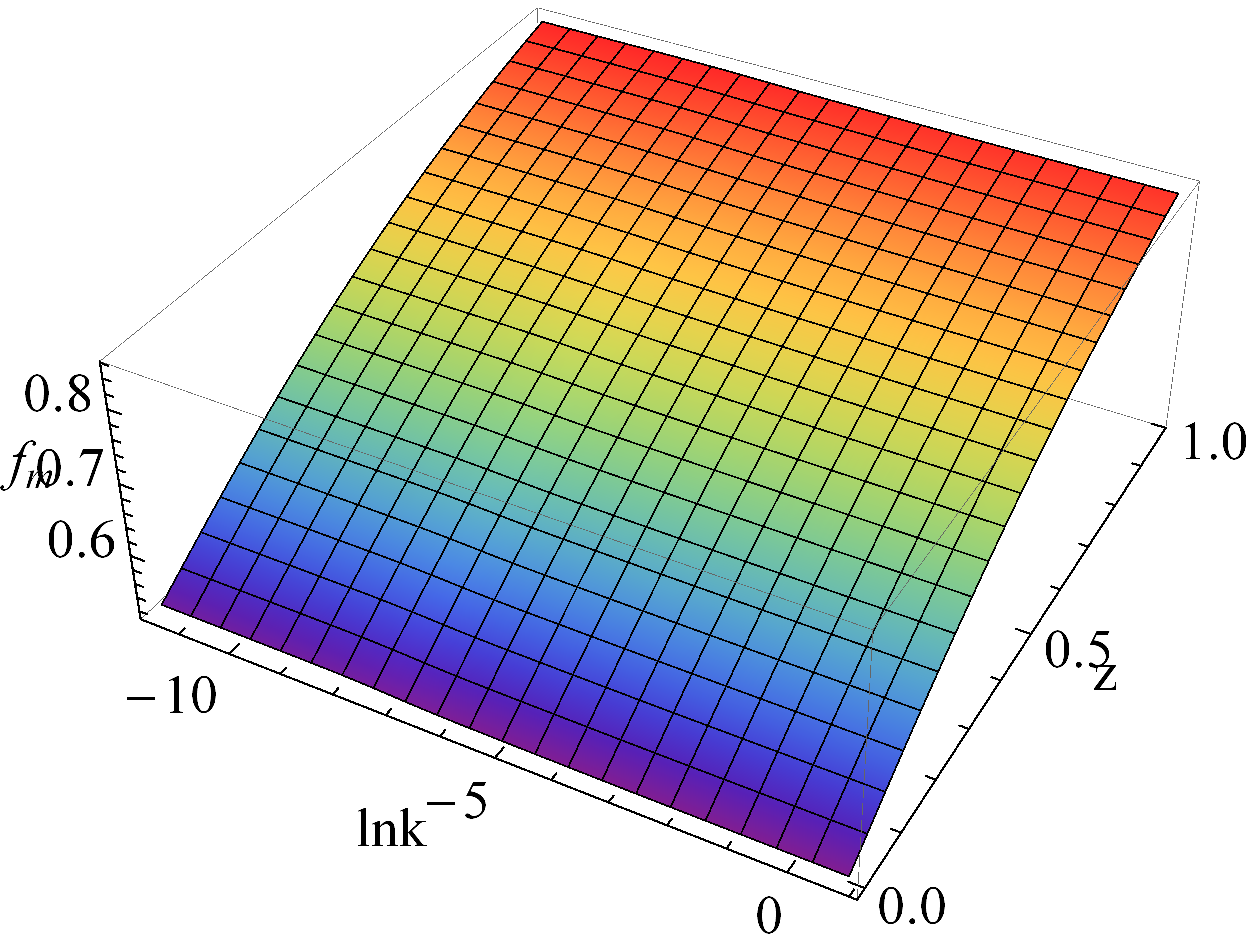}
\includegraphics[width=8.8cm,height=6.8cm]{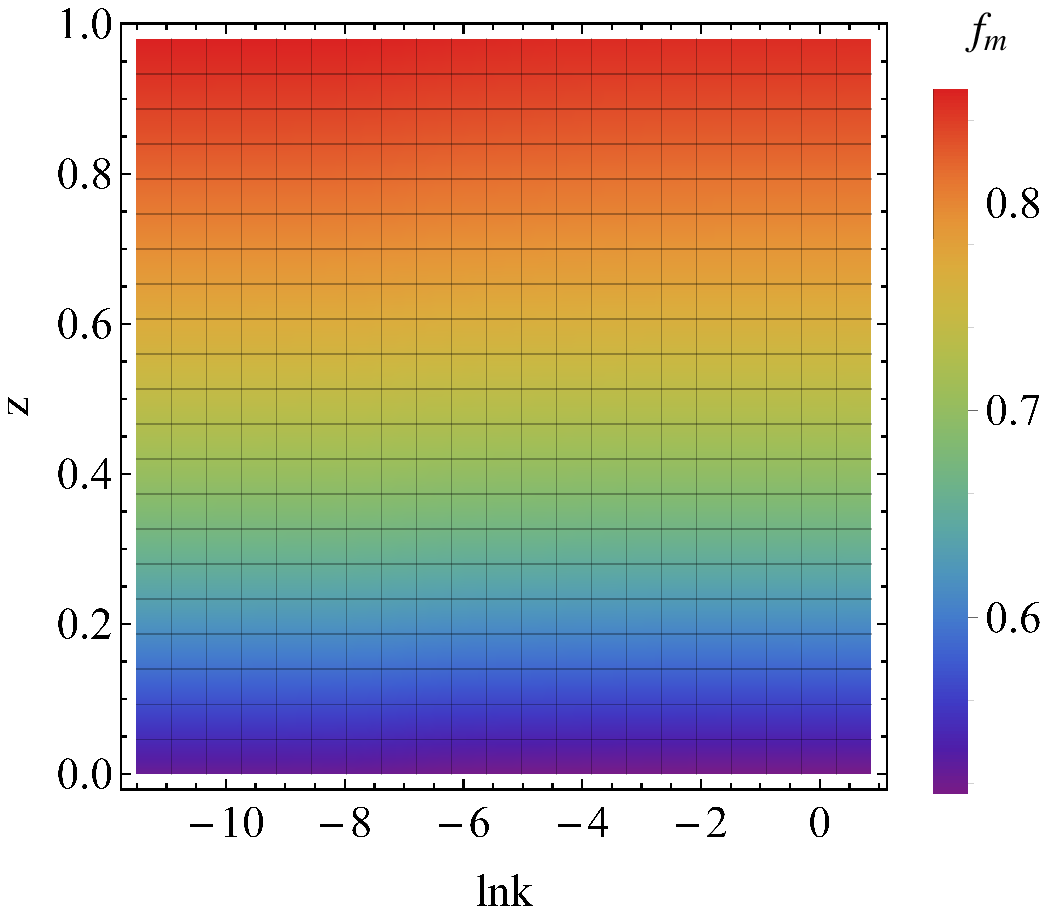}
  \caption{The three-dimensional plots of $\ln k$ ($k$ is the wavenumber), $z$ (redshift), $f_m$ (growth rate of matter). Here, $\xi_x$ and the other relevant parameters are fixed with the mean values as shown in the fifth column of Table \ref{tab:results-mean-pos}.}
  \label{fig:growth-kzf}
\end{figure}

Here, it is necessary to explain how to modify the \textbf{CAMB} code \cite{ref:camb} and \textbf{CosmoMC} package \cite{ref:cosmomc-Lewis2002}. We not only modify the \textbf{CAMB} code \cite{ref:camb} based on the continuity and Euler equations about the dark sectors, but also add some codes to calculate the density perturbations of the matter via $\delta_m=(\rho_c\delta_c+\rho_b\delta_b)/(\rho_c+\rho_b)$. In the light of $f_m=d\ln\delta_m/d\ln a=\delta'_m/(\mathcal{H}\delta_m)$, we could calculate the theoretical values of growth rate for matter, and put them into a three dimensional table about the wavenumber $k$, redshift $z$, and growth rate $f_m$. When $\xi_x$ and the other relevant parameters are fixed with the mean values, we present the three-dimensional plots of $\ln k$, $z$, and $f_m$ in Fig. \ref{fig:growth-kzf}. With decreasing the values of $z$, the growth rate $f_m$ is decreased. Besides, when $z$ is fixed, from Fig. \ref{fig:growth-kzf}, it is easy to see that the growth rate is scarcely dependent on the scale. Therefore, in the theoretical frame of general relativity, we could consider that the linear growth is scale independent \cite{ref:fsigma8total-Samushia2013,ref:fsigma84-Reid2012}. In order to adopt the RSD measurement, we add a new module \textbf{CosmoMC} package \cite{ref:cosmomc-Lewis2002} to import $f_m$ from CAMB which could be used to calculate the theoretical values of $f\sigma_8(z)$ at ten different redshifts. For constraining the other cosmological models with the RSD analysis, please see Refs. \cite{ref:fsigma8-HDE-Xu2013,ref:fsigma8andPlanck-MG-Xu2013,ref:Xu2013-DGP,ref:Yang2013-1,ref:Yang2013-2,ref:Yang2014-ux}.

\begin{figure}[!htbp]
\includegraphics[width=13cm,height=9cm]{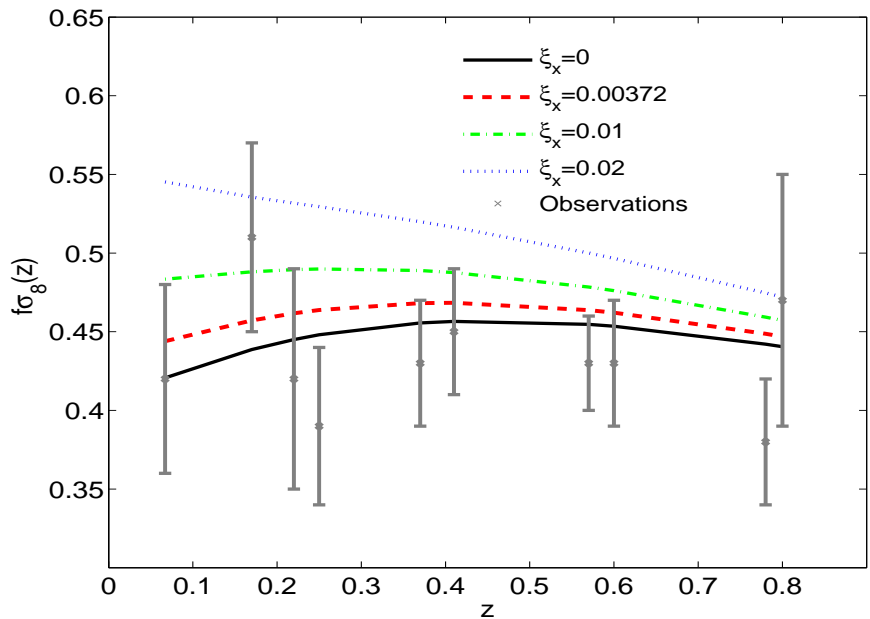}
  \caption{The fitting evolutions of $f\sigma_8(z)$ about the redshift $z$ for varied interaction rate $\xi_x$. The black solid, red thick dashed, green dotted-dashed, and blue dotted lines are for $\xi_x=0, 0.00372, 0.01$, and $0.02$, respectively; the gray error bars denote the observations of $f\sigma_8(z)$ at different redshifts are listed in Table \ref{tab:fsigma8data}; the other relevant parameters are fixed with the mean values as shown in the fifth column of Table \ref{tab:results-mean-pos}.}
  \label{fig:fsigma8-pos}
\end{figure}

Furthermore, in order to investigate the effects of interaction rate $\xi_x$ to $f\sigma_8(z)$, we fix the relevant mean values of our constraint results in Table \ref{tab:results-mean-pos}, but keep the model parameter $\xi_x$ varying in a range. At ten different redshifts, we derive the theoretical values of the growth function from the new module in the modified CosmoMC package. When $\xi_x$ is fixed on a value, We fit the ten theoretical data points (z, $f\sigma_8(z)$) and plot the evolution curves of $f\sigma_8(z)$ in Fig. \ref{fig:fsigma8-pos}.
Here, we could make a qualitative analysis on the relation between varied $\xi_x$ and changed $f\sigma_8(z)$. Positive interaction rate denotes a transfer of energy from dark matter to dark energy, with fixed $\Omega_c$ today, the dark matter energy density would be greater in the past than the uncoupled case. A larger proportion of dark matter naturally leads to more structure growth (as is shown in Fig. \ref{fig:Dc-fc}) and the increase of present matter power spectra (as is shown in Fig. \ref{fig:Mpower-pos}), which are correspondingly the larger growth rate and the higher $\sigma_8$ ($\sigma_8$ could be obtained by the integration with regard to the matter power spectra \cite{ref:fsigma8-DE-Song2009,ref:Percival2009}). Therefore, the values of $f\sigma_8(z)$ are enhanced than the uncoupled case, and the amplitude of enhancement becomes obvious with raising the values of $\xi_x$. Besides, from Eqs. (\ref{eq:thetac-prime2-ide2-mod}), (\ref{eq:Heff}), and (\ref{eq:Geff}), we also could know why the changed amplitude of $f\sigma_8(z)$ becomes large with reducing the redshift. For fixed $\xi_x$, at the higher redshift, the component of dark energy is subdominant, the modified Hubble friction term and source term are trivial, which would slightly affect the evolutions of growth rate and $\sigma_8$. Nonetheless, at the lower redshift, the dark energy gradually dominates the late Universe, the modified $H_{eff}$ and $G_{eff}$ would significantly increase the cosmic structure growth, which could bring about more obvious enhancement of $f\sigma_8(z)$.
Particularly, it is easy to see that the case of $\xi_x=0.00372$ (corresponds the IwCDM model with mean value) and that of $\xi_x=0$ (corresponds to the uncoupled wCDM model) are significantly distinguishing from the evolution curves of $f\sigma_8(z)$, which is different from the evolutions of CMB temperature and matter power spectra. It means that, to some extent, the RSD test could break the possible degeneracy between the IwCDM model and the uncoupled wCDM model.

\subsection{Cosmological constraint results}

In our numerical calculations, the total likelihood is calculated by $\mathcal{L}\propto e^{-\chi^2/2}$, where $\chi^2$ can be constructed as
\begin{eqnarray}
\chi^2_{total}=\chi^2_{CMB}+\chi^2_{BAO}+\chi^2_{SN}+\chi^2_{RSD}.
\label{eq:chi22}
\end{eqnarray}
where the four terms in right side of this equation, respectively, denote the contribution from CMB, BAO, SN, and RSD data sets. The used data sets for our MCMC likelihood analysis are listed in Table \ref{tab:alldata}. Some detailed descriptions about the observed data sets have been shown in Appendix C of this paper.

\begin{table}
\begin{center}
\begin{tabular}{ccc}
\hline\hline Data names & Data descriptions and references \\ \hline
CMB & $l\in[50,2500]$ temperature likelihood from \textit{Planck} \cite{ref:Planck2013-params} \\
$-$ & up to $l=49$ temperature likelihood from \textit{Planck} \cite{ref:Planck2013-params} \\
$-$ & up to $l=32$ polarization likelihood from WMAP9 \cite{ref:WMAP9} \\
BAO & $r_s/D_V(z=0.106)=0.336\pm0.015$ \cite{ref:BAO-1}\\
$-$ & $r_s/D_V(z=0.35)=0.1126\pm0.0022$ \cite{ref:BAO-2}\\
$-$ & $r_s/D_V(z=0.57)=0.0732\pm0.0012$ \cite{ref:BAO-3}\\
SNIa & SNLS3 data from SiFTO and SALT2 \cite{ref:SNLS3-1,ref:SNLS3-2,ref:SNLS3-3}\\
RSD & ten $f\sigma_8(z)$ data points from Table \ref{tab:fsigma8data}\\
\hline\hline
\end{tabular}
\caption{The used data sets for our MCMC likelihood analysis on the coupled dark energy model, where $l$ is the multipole number of power spectra, WMAP9 is the abbreviation of nine-year Wilkinson Microwave Anisotropy Probe, and SNLS3 is the abbreviation of three-year Supernova Legacy Survey.}
\label{tab:alldata}
\end{center}
\end{table}

For the IwCDM model, we consider the eight-dimensional parameter space which reads
\begin{eqnarray}
P\equiv\{\Omega_bh^2, \Omega_{c}h^2, \Theta_S, \tau, w_x, \xi_x, n_s, log[10^{10}A_S]\},
\label{eq:parameter_space}
\end{eqnarray}
where $\Omega_bh^2$ and $\Omega_{c}h^2$, respectively, stand for the density of the baryons and dark matter, $\Theta_S=100\theta_{MC}$ refers to the ratio of sound horizon and angular diameter distance, $\tau$ indicates the optical depth, $w_x$ is the EoS of dark energy, $\xi_x$ is the interaction rate between the dark sectors, $n_s$ is the scalar spectral index, and $A_s$ represents the amplitude of the initial power spectrum. The priors to the basic model parameters are listed in the second column of Table \ref{tab:results-mean-pos}.
Here, the pivot scale of the initial scalar power spectrum $k_{s0}=0.05Mpc^{-1}$ is used. Then, based on the MCMC method, we perform a global fitting for the interacting model with $Q^{\mu}_A\parallel u^{\mu}_c$ when the model parameters satisfy $\xi_x>0$ and $(1+w_x)>0$. Here, we choose $c^2_{sx}=1$ which could avoid the unphysical sound speed \cite{ref:Valiviita2008,ref:Majerotto2010,ref:Clemson2012}.

\begin{figure}[!htbp]
\includegraphics[width=20cm,height=15cm]{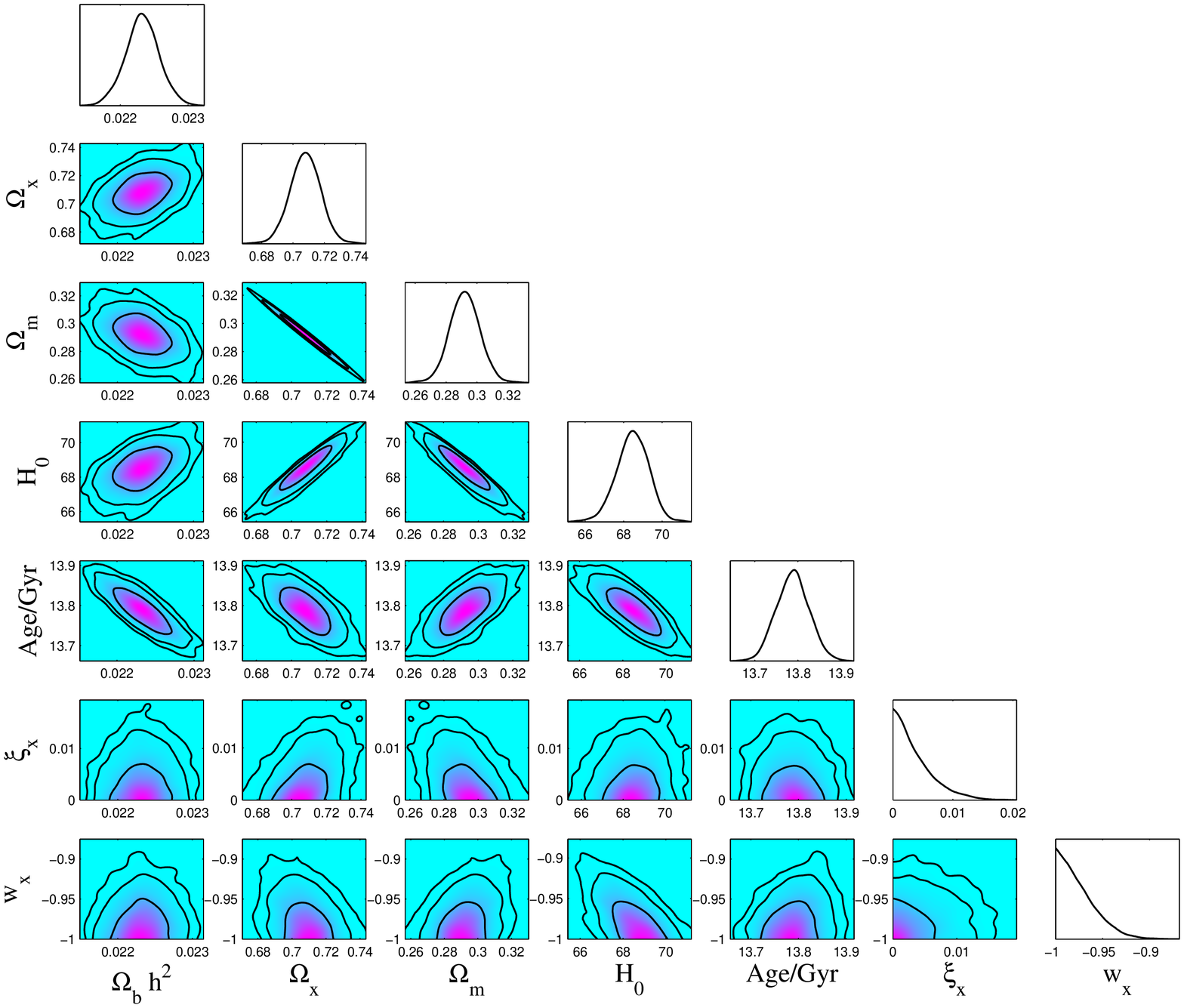}
  \caption{The 1D marginalized distributions on individual parameters and 2D contours with 68\% C.L., 95 \% C.L., and 99.7\% C.L. between each other using the combination of the observed data points from the CMB from \textit{Planck} + WMAP9, BAO, SN, and RSD data sets.}
  \label{fig:contour-pos}
\end{figure}

After running eight chains in parallel on the computer, the constraint results for the IwCDM model are, respectively, presented in the fifth and sixth columns of Table \ref{tab:results-mean-pos}. We show the one-dimensional (1D) marginalized distributions of parameters and two-dimensional (2D) contours with $68\%$ C.L., $95\%$ C.L., and $99.7\%$ C.L. in Figs. \ref{fig:contour-pos}.
We anticipate that the large scale structure test will give a tighter constraint on the parameter space than before. In order to compare with the constraint without RSD data, we also constrain the IwCDM model without the $f\sigma_8(z)$ data set, the results are shown in the third and fourth columns of Table \ref{tab:results-mean-pos}.

Here, we pay attention to the constraint result of the interaction rate. In the third column of Table \ref{tab:results-mean-pos}, we find the interaction rate $\xi_x=0.209^{+0.0711}_{-0.0403}$ in 1$\sigma$ region. Some similar constraint results have been presented in the previous papers. Before \textit{Planck} data, $Q=\Gamma_x\rho_x$ (belongs to the interacting model (III)) was considered in Ref. \cite{ref:Clemson2012}, the interacting dark energy with a constant EoS has been constrained by CMB from WMAP7 \cite{ref:WMAP7}, BAO \cite{ref:BAO-Clemson2012}, HST (Hubble Space Telescope) \cite{ref:HST-Clemson2012} and SN from SDSS \cite{ref:SNIa-Clenmson2012}, the results of $Q^{\mu}_A\parallel u^{\mu}_c$ showed that the best-fit value of interaction rate was $\Gamma_x/H_0=0.366$. After \textit{Planck} data, in Ref. \cite{ref:Salvatelli2013}, the perturbed expansion rate of the Universe and the interacting form $Q=H\xi_x\rho_x$ was considered, this interacting model has been tested by CMB from \textit{Planck} + WMAP9 \cite{ref:Planck2013-params,ref:WMAP9}, BAO \cite{ref:BAO-1,ref:BAO-2,ref:BAO-3} and HST \cite{ref:HST-Salvatelli2013}. The constraint results from CMB and BAO presented that the mean values of interaction rate were $\xi_x=-0.61^{+0.12}_{-0.25}$ from CMB and BAO measurements, and $\xi_x=-0.67^{+0.086}_{-0.17}$ from CMB and HST tests (the minus is from the background evolution equations of dark matter and dark energy).


In brief summary, the geometry tests which mainly include CMB, BAO, SN, and HST slightly favor the interaction between dark matter and dark energy. Meanwhile, the growth rate of dark matter perturbations possibly rules out large interaction rate which was pointed out in Ref. \cite{ref:Clemson2012}. Instead of the case without RSD data, the large scale structure information evidently influences the expansion history of the Universe and the evolution of matter density perturbations, the parameter space of the interacting model is greatly improved. As expected, from the fifth column of Table \ref{tab:results-mean-pos}, we find the recently cosmic observations indeed favor small interaction rate $\xi_x=0.00372_{-0.00372}^{+0.000768}$ after the RSD measurement is added. To some extent, the $f\sigma_8(z)$ test could rule out large interaction rate.

Furthermore, based on the same observed data sets (CMB from \textit{Planck} + WMAP9, BAO, SN and RSD), the IwCDM model has another two parameters $w_x$ and $\xi_x$ which give rise to the difference of the minimum $\chi^2$ with the $\Lambda$CDM model, $\Delta\chi^2_{min}=2.819$.

\begingroup
\squeezetable
\begin{center}
\begin{table}
\begin{tabular}{cccccccc}
\hline\hline Parameters & Priors & IwCDM without RSD & Best fit & IwCDM with RSD & Best fit & $\Lambda$CDM with RSD & Best fit \\ \hline
$\Omega_bh^2$&[0.005,0.1]&$0.0220_{-0.000242-0.000479-0.000620}^{+0.000244+0.000503+0.000663}$&$0.0221$&$ 0.0223_{-0.000240-0.000490-0.000613}^{+0.000233+0.000490+0.000642}$&$0.0223$&$0.0223_{-0.000245-0.000461- 0.000602}^{+0.000245+0.000495+0.000642}$&$0.0225$
\\
$\Omega_ch^2$&[0.01,0.99]&$0.0390_{-0.0374-0.0380-0.0380}^{+0.0115+0.0417+0.0567}$&$0.0464$&$0.114_{- 0.00171-0.00405-0.00602}^{+0.00217+0.00385+0.00450}$&$0.115$&$0.116_{-0.00145-0.00282-0.00365}^{+0.00144+ 0.00286+0.00376}$&$0.115$
\\
$100\theta_{MC}$&[0.5,10]&$1.0464_{-0.00209-0.00362-0.00445}^{+0.00217+0.00374+0.00432}$&$1.0455$&$ 1.0416_{-0.000573-0.00113-0.00145}^{+0.000570+0.00111+0.00139}$&$1.0413$&$1.0407_{-0.000551-0.00105- 0.00138}^{+0.000543+0.00105+0.00143}$&$1.0408$
\\
$\tau$&[0.01,0.8]&$0.0882_{-0.0136-0.0239-0.0310}^{+0.0122+0.0257+0.0348}$&$0.0828$&$0.0862_{-0.0122- 0.0226-0.0305}^{+0.0120+0.0239+0.0337}$&$0.0831$&$0.0860_{-0.0128-0.0229-0.0293}^{+0.0117+0.0250+0.0325}$& $0.0788$
\\
$\xi_x$&[0,1]&$0.209_{-0.0403-0.113-0.153}^{+0.0711+0.0969+0.110}$&$0.203$&$0.00372_{-0.00372-0.00372- 0.00372}^{+0.000768+0.00655+0.0102}$&$0.00328$&$---$&$-$
\\
$w_x$&[-1,0]&$-0.940_{-0.0599-0.0599-0.0599}^{+0.0158+0.0817+0.115}$&$-0.998$&$-0.975_{-0.0246-0.0246- 0.0246}^{+0.00581+0.0382+0.0601}$&$-0.995$&$---$&$-$
\\
$n_s$&[0.5,1.5]&$0.967_{-0.00566-0.0109-0.0142}^{+0.00564+0.0112+0.0144}$&$0.967$&$0.977_{-0.00550-0.0107- 0.0139}^{+0.00550+0.0109+0.0145}$&$0.975$&$0.969_{-0.00542-0.0109-0.0148}^{+0.00538+0.0107+0.0146}$&$ 0.972$
\\
${\rm{ln}}(10^{10}A_s)$&[2.4,4]&$3.0974_{-0.0244-0.0475-0.0622}^{+0.0244+0.0497+0.0701}$&$3.0910$&$ 3.0802_{-0.0232-0.0441-0.0603}^{+0.0229+0.0467+0.0642}$&$3.0784$&$3.0719_{-0.0232-0.0444-0.0565}^{+0.0232+ 0.0488+0.0630}$&$3.0559$
\\
\hline
$\Omega_x$&$-$&$0.877_{-0.0324-0.0904-0.125}^{+0.0668+0.0783+0.0783}$&$0.866$&$0.708_{-0.00940-0.0187- 0.0273}^{+0.00929+0.0187+0.0274}$&$0.705$&$0.710_{-0.00819-0.0167-0.0224}^{+0.00815+0.0158+0.0120}$&$ 0.713$
\\
$\Omega_m$&$-$&$0.123_{-0.0668-0.0783-0.0783}^{+0.0324+0.0904+0.125}$&$0.134$&$0.292_{-0.00929-0.0187- 0.0274}^{+0.00940+0.0188+0.0273}$&$0.295$&$0.290_{-0.00815-0.0158-0.0199}^{+0.00819+0.0167+0.0224}$&$ 0.287$
\\
$\sigma_8$&$-$&$---$&$-$&$0.804_{-0.0113-0.0244-0.0332}^{+0.0121+0.0234+0.0323}$&$0.812$&$0.810_{-0.0109- 0.0190-0.0249}^{+0.00992+0.0201+0.0263}$&$0.802$
\\
$z_{re}$&$-$&$10.974_{-1.0935-2.174-2.863}^{+1.0891+2.136+2.886}$&$10.512$&$10.583_{-1.0354-1.993- 2.735}^{+1.0162+2.0164+2.694}$&$10.362$&$10.570_{-1.0183-2.0293-2.670}^{+1.0229+2.0821+2.635}$&$9.904$
\\
$H_0$&$-$&$71.0830_{-1.218-2.412-3.137}^{+1.297+2.371+3.0613}$&$71.932$&$68.462_{-0.759-1.657-2.385}^{+ 0.887+1.536+2.181}$&$68.479$&$69.130_{-0.665-1.315-1.745}^{+0.677+1.336+1.692}$&$69.456$
\\
${\rm{Age}}/{\rm{Gyr}}$&$-$&$13.774_{-0.0357-0.0714-0.0953}^{+0.0353+0.0703+0.0933}$&$13.752$&$13.788_{- 0.0381-0.0705-0.0952}^{+0.0375+0.0737+0.0968}$&$13.791$&$13.756_{-0.0344-0.0709-0.0932}^{+0.0376+0.0683+ 0.0897}$&$13.737$
\\
\hline\hline
\end{tabular}
\caption{The mean values with $1,2,3\sigma$ errors and the best fit values of the parameters for the IwCDM model and the $\Lambda$CDM model, where CMB from \textit{Planck} + WMAP9, BAO, SN, with or without RSD data sets have been used.}
\label{tab:results-mean-pos}
\end{table}
\end{center}
\endgroup

\section{SUMMARY}

In this paper, we considered a type of interaction which was relative to the expansion rate $H$ of the Universe. When the interaction was introduced, the effective EoS of dark energy brought about the deviation from the uncoupled case. In a general gauge, via introducing the parameter of 'choosing the momentum transfer $b$ for A fluid, we obtained two sets of different perturbation equations in the rest frame of dark matter or dark energy. Furthermore, in the synchronous gauge, based on the interaction form $Q=3H\xi_x\rho_x$ whose perturbation equations were stable in most cases, the continuity and Euler equations were gained for the IwCDM model. According to the density perturbations of dark matter and baryons, we added a module to calculate the theoretical values of $f\sigma_8(z)$ which could be used to constrain the IwCDM model. In the aspect of theoretical predictions, we have plotted the effects of the varied interaction rate on CMB power spectra and matter power spectra. Then, we have shown the modified growth of structure with the varied interaction rate, and presented the deviations from standard model of the effective expansion rate and effective gravitational constant for the density perturbations of dark matter. We also plotted the evolution curves of $f\sigma_8(z)$. From the panel of $f\sigma_8(z)$, we could clearly distinguish from the IwCDM model with mean value to the uncoupled wCDM model, meanwhile, the CMB and matter power spectra could not make it. It meant that, to some extent, $f\sigma_8(z)$ could break the possible degeneracy of the cosmological models. Based on the MCMC method, we constrained the interacting model by CMB from \textit{Planck} + WMAP9, BAO, SN, and the RSD test. After adding the measurement of large scale structure information, we received a tighter constraint on the model parameters than the case without RSD data set. Instead of the case without RSD data, the large scale structure information evidently influences the expansion history of the Universe and the evolution of matter density perturbations, the parameter space of the interacting model is greatly improved. Moreover, we found the interaction rate in 3$\sigma$ regions: $\xi_x=0.00372_{-0.00372- 0.00372-0.00372}^{+0.000768+0.00655+0.0102}$. The currently available cosmic observations favor small interaction rate between the dark sectors, at the same time, the $f\sigma_8(z)$ test could rule out large interaction rate in 1$\sigma$ region.

For the interacting model (III), we have constrained the IwCDM model for the case of $Q^{\mu}_A\parallel u^{\mu}_c$, next, the other case of interacting dark energy $Q^{\mu}_A\parallel u^{\mu}_x$ will be constrained by the recently cosmic observations. Moreover, if one would consider the perturbations about expansion rate of the Universe \cite{ref:Gavela2010}, the continuity and Euler equations for the IwCDM model are shown in Appendix B of this paper. Besides, we would continue to study the interacting models (I) and (II) and try to constrain the interaction rate. Last but the most important is that we will go on exploring the effects on the cosmological constraints from the large scale structure information.

\acknowledgements{L. Xu's work is supported in part by NSFC under the Grants No. 11275035 and "the Fundamental Research Funds for the Central Universities" under the Grants No. DUT13LK01.}

\appendix
\section{Verifying the perturbation equations}

For an application example for Eqs. (\ref{eq:general-deltax-b},\ref{eq:general-deltac-b},\ref{eq:general-thetax-b},\ref{eq:general-thetac-b}), if we follow Ref. \cite{ref:Clemson2012} and take the interaction $Q_x=-Q_c=Q=\Gamma_x\rho_x$, so $\delta Q_x=-\delta Q_c=\Gamma_x\rho_x\delta_x$. Moreover, in order to avoid the unphysical sound speed, we choose $c^2_{sx}=1$ \cite{ref:Valiviita2008,ref:Majerotto2010,ref:Clemson2012}. Under these conditions, we could obtain the continuity and Euler equations which are compatible with Eqs. (32-37) in Ref. \cite{ref:Clemson2012}
\begin{eqnarray}
\delta'_x+3\mathcal{H}(1-w_x)\delta_x
+9\mathcal{H}^2(1-w_x^2)\frac{\theta_x}{k^2}
+(1+w_x)\theta_x-3(1+w_x)\psi'+(1+w_x)k^2(B-E')
\nonumber \\
=a\Gamma_x\left[\phi+3\mathcal{H}(1-w_x)\frac{\theta_x}{k^2}\right],
\label{eq:Clemson-deltax-b}
\end{eqnarray}
\begin{eqnarray}
\delta'_c+\theta_c-3\psi'+k^2(B-E')
=a\Gamma_x\frac{\rho_x}{\rho_c}(\delta_c-\delta_x-\phi),
\label{eq:Clemson-deltac-b}
\end{eqnarray}
\begin{eqnarray}
\theta'_x-2\mathcal{H}\theta_x-\frac{k^2\delta_x}{(1+w_x)}-k^2\phi
=\frac{a\Gamma_x}{(1+w_x)}[b\theta_c-(1+b)\theta_x],
\label{eq:Clemson-thetax-b}
\end{eqnarray}
\begin{eqnarray}
\theta'_c+\mathcal{H}\theta_c-k^2\phi
=a\Gamma_x\frac{\rho_x}{\rho_c}(1-b)(\theta_c-\theta_x),
\label{eq:Clemson-thetac-b}
\end{eqnarray}
where
\[b = \left\{ \begin{array}{l}
 1,~~for~~Q^{\mu}_A~\parallel~u^{\mu}_c, \\
 0,~~for~~Q^{\mu}_A~\parallel~u^{\mu}_x, \\
 \end{array} \right.\]

\section{Verifying the perturbation equations when the expansion rate of the Universe is perturbed}

If one considers the expansion rate of the Universe is perturbed in the light of Ref. \cite{ref:Gavela2010}, $\tilde{H}=H+\delta H$. When the interacting form is taken as $Q_x=-Q_c=H\xi_x\rho_x$, one could obtain the continuity and Euler equations of dark energy and dark matter
\begin{eqnarray}
\delta Q_x=-\delta Q_c
=3H\xi_x\rho_x\left(\frac{\delta H}{H}+\frac{\delta\rho_x}{\rho_x}\right)
=3H\xi_x\rho_x(\mathcal{K}+\delta_x),
\label{eq:deltaQ-ywq-dH}
\end{eqnarray}
where $\mathcal{K}=\delta H/H$, according to Eqs. (\ref{eq:general-deltax-b},\ref{eq:general-deltac-b},\ref{eq:general-thetax-b},\ref{eq:general-thetac-b}), the continuity and Euler equations for the IwCDM model read

\begin{eqnarray}
\delta'_x+3\mathcal{H}(c^2_{sx}-w_x)\delta_x
+9\mathcal{H}^2(1+w_x)(c^2_{sx}-w_x)\frac{\theta_x}{k^2}
+(1+w_x)\theta_x-3(1+w_x)\psi'+(1+w_x)k^2(B-E')
\nonumber \\
=\mathcal{H}\xi_x\phi
+3\mathcal{H}^2(c^2_{sx}-w_x)\xi_x\frac{\theta_x}{k^2}
+\mathcal{H}\xi_x\mathcal{K},
\label{eq:Clemson-deltax-b-dH}
\end{eqnarray}
\begin{eqnarray}
\delta'_c+\theta_c-3\psi'+k^2(B-E')
=\mathcal{H}\xi_x\frac{\rho_x}{\rho_c}(\delta_c-\delta_x)
-\mathcal{H}\xi_x\frac{\rho_x}{\rho_c}\phi
-\mathcal{H}\xi_x\frac{\rho_x}{\rho_c}\mathcal{K},
\label{eq:Clemson-deltac-b-dH}
\end{eqnarray}
\begin{eqnarray}
\theta'_x+\mathcal{H}(1-3c^2_{sx})\theta_x-\frac{c^2_{sx}}{(1+w_x)}k^2\delta_x-k^2\phi
=\frac{\mathcal{H}\xi_x}{1+w_x}
[b(\theta_c-\theta_x)-c^2_{sx}\theta_x],
\label{eq:Clemson-thetax-b-dH}
\end{eqnarray}
\begin{eqnarray}
\theta'_c+\mathcal{H}\theta_x-k^2\phi
=\mathcal{H}\xi_x\frac{\rho_x}{\rho_c}(1-b)(\theta_c-\theta_x),
\label{eq:Clemson-thetac-b-dH}
\end{eqnarray}

Furthermore, in the synchronous gauge, $\mathcal{K}=\delta H/H=(\theta+h'/2)/(3\mathcal{H})$ \cite{ref:Gavela2010}, so the continuity and Euler equations become

\begin{eqnarray}
\delta'_x+(1+w_x)\left(\theta_x+\frac{h'}{2}\right)+3\mathcal{H}(c^2_{sx}-w_x)\delta_x
+9\mathcal{H}^2(c^2_{sx}-w_x)(1+w_x)\frac{\theta_x}{k^2}
\nonumber \\
=3\mathcal{H}^2(c^2_{sx}-w_x)\xi_x\frac{\theta_x}{k^2}
+\frac{\xi_x}{3}\left(\theta+\frac{h'}{2}\right),
\label{eq:deltax-prime-b-dH-syn}
\end{eqnarray}
\begin{eqnarray}
\delta'_c+\theta_c+\frac{h'}{2}
=\mathcal{H}\xi_x\frac{\rho_x}{\rho_c}(\delta_c-\delta_x)
-\frac{\xi_x}{3}\frac{\rho_x}{\rho_c}\left(\theta+\frac{h'}{2}\right),
\label{eq:deltac-prime-b-dH-syn}
\end{eqnarray}
\begin{eqnarray}
\theta'_x+\mathcal{H}(1-3c^2_{sx})\theta_x-\frac{c^2_{sx}}{1+w_x}k^2\delta_x
=\frac{\mathcal{H}\xi_x}{1+w_x}
[b(\theta_c-\theta_x)-c^2_{sx}\theta_x],
\label{eq:thetax-prime-b-dH-syn}
\end{eqnarray}
\begin{eqnarray}
\theta'_c+\mathcal{H}\theta_c
=\mathcal{H}\xi_x\frac{\rho_x}{\rho_c}(1-b)(\theta_c-\theta_x),
\label{eq:thetac-prime-b-dH-syn}
\end{eqnarray}

where $(\rho+p)v=\sum\limits_A (\rho_A+p_A)v_A$ \cite{ref:Valiviita2008,ref:Gavela2010} and $\theta_A=-k^2(v_A+B)$ \cite{ref:Valiviita2008,ref:Ma1995}.

In the case of $Q^{\mu}_A\parallel u^{\mu}_c$ (b=1), when $c^2_{sx}=1$ \cite{ref:Valiviita2008,ref:Majerotto2010,ref:Clemson2012}, we could obtain the continuity and Euler equations which are compatible with Eqs. (4.1-4.4) in Ref. \cite{ref:Gavela2010}. (Here, $\xi_x=-\xi$ \cite{ref:Gavela2010} because the background evolution equations of dark matter and dark energy are different between the two works.) Moreover, we generalize the continuity and Euler equations into the case of $Q^{\mu}_A\parallel u^{\mu}_x$ (b=0).

\section{Observed data sets}

For the BAO data set, we use the measured ratio of $r_s/D_v$ as a 'standard ruler', $r_s$ is the comoving sound horizon at the baryon drag epoch, $D_v$ is the effective distance which is determined by the angular diameter distance $D_A$ and Hubble parameter $H$ \cite{ref:Eisenstein2005,ref:Percival2007}
\begin{eqnarray}
D_v(z)=\left[(1+z)^2D_A(a)^2\frac{z}{H(z)}\right]^{1/3}.
\label{eq:Dv}
\end{eqnarray}

At three different redshifts, $r_s(z_d)/D_V(z=0.106)=0.336\pm0.015$ is from 6-degree Field Galaxy Redshift Survey (6dFGRS) data \cite{ref:BAO-1}, $r_s(z_d)/D_V(z=0.35)=0.1126\pm0.0022$ comes from Sloan Digital
Sky Survey Data Release 7 (SDSS DR7) data \cite{ref:BAO-2}, and $r_s(z_d)/D_V(z=0.57)=0.0732\pm0.0012$ is from SDSS DR9 \cite{ref:BAO-3}. So, the likelihood for BAO reads

\begin{eqnarray}
\chi^2_{BAO}&=&\chi^2_{6dF}+\chi^2_{DR7}+\chi^2_{DR9} \\ \nonumber
&=&\frac{\left[\left(r_s(z_d)/D_V(0.106)\right)_{th}-0.336\right]^2}{0.015^2}
+\frac{\left[\left(r_s(z_d)/D_V(0.35)\right)_{th}-0.1126\right]^2}{0.0022^2}
+\frac{\left[\left(r_s(z_d)/D_V(0.57)\right)_{th}-0.0732\right]^2}{0.0012^2}.
\label{eq:BAO}
\end{eqnarray}

For the SN data set, we use the SNLS3 data, which consists of 472 SN calibrated by SiFTO and SALT2 \cite{ref:SNLS3-1,ref:SNLS3-2,ref:SNLS3-3}. The likelihood for this sample is constructed as \cite{ref:SNLS3-2,ref:SNLS3-3}
\begin{eqnarray}
\chi^2_{SN}=(\overrightarrow{m}_B-\overrightarrow{m}_B^{model})^T C_{SN}^{-1}
(\overrightarrow{m}_B-\overrightarrow{m}_B^{model}).
\label{eq:SNLS}
\end{eqnarray}
where $\overrightarrow{m}_B$ is the vector of effective absolute magnitudes and $C_{SN}^{-1}$ is the sum of non-sparse covariance matrices of quantifying statistical and systematic errors \cite{ref:SNLS3-2}. The expected apparent magnitudes of cosmological model are given by \cite{ref:SNLS3-2,ref:SNLS3-3}
\begin{eqnarray}
m_B^{model}=5log_{10}\mathcal{D}_L(z_{hel},z_{cmb},w_x,\Omega_m,\Omega_{x})
-\alpha(s-1)+\beta\mathcal{C}+\mathcal{M}_B,
\label{eq:mb-model}
\end{eqnarray}
where $\mathcal{D}_L$ is the Hubble-constant free luminosity distance, $z_{cmb}$ and $z_{hel}$ are the CMB frame and heliocentric redshifts of the SN, $s$ is the stretch (a measure of the shape of the SN light curve), and $\mathcal{C}$ is color measure for the SN. $\alpha$ and $\beta$ are nuisance parameters. $\mathcal{M}_B$ is another nuisance parameter which absorbs the Hubble constant. As in Ref. \cite{ref:SNLS3-3}, one could express values of the parameter $\mathcal{M}_B$ in term of an effective absolute magnitude, $m_B=\mathcal{M}_B-5log_{10} (c/H_0)-25$.

After \textit{Planck} data, the CMB data which includes two main parts: one is the high-l temperature likelihood (\textit{CAMSpec}) up to a maximum multipole number of $l_{max}=2500$ from $l=50$ \cite{ref:Planck2013-params}; the other is the low-l temperature likelihood up to $l=49$ \cite{ref:Planck2013-params} and the low-l polarization likelihood up to $l=32$ from nine-year WMAP data \cite{ref:WMAP9}.

The likelihood of RSD measurement is given by
\begin{eqnarray}
\chi^2_{RSD}=\sum\frac{[f\sigma_8(z_i)_{th}-f\sigma_8(z_i)_{obs}]}{\sigma_{i}^2}.
\label{eq:RSD}
\end{eqnarray}

Therefore, the total likelihood is calculated by $\mathcal{L}\propto e^{-\chi^2/2}$, where $\chi^2$ can be constructed as
\begin{eqnarray}
\chi^2_{total}=\chi^2_{BAO}+\chi^2_{SN}+\chi^2_{CMB}+\chi^2_{RSD}.
\label{eq:chi22}
\end{eqnarray}


\begin{thebibliography}{*}

\bibitem{ref:Planck2013} P.A.R. Ade, \textit{et al.}, Planck Collaboration, [arXiv:1303.5062].
\bibitem{ref:Planck2013-CMB} P.A.R. Ade, \textit{et al.}, Planck Collaboration, [arXiv:1303.5075].
\bibitem{ref:Planck2013-params} P.A.R. Ade, \textit{et al.}, Planck Collaboration, [arXiv:1303.5076].
\bibitem{ref:Planck2013-download} http://pla.esac.esa.int/pla/aio/planckProducts.html.

\bibitem{ref:WMAP9} G. Hinshaw, \textit{et al.}, Astrophys. J. Suppl. \text{208}, 19 (2013). [arXiv:1212.5226]

\bibitem{ref:Zlatev1999} I. Zlatev, L. Wang, and P.J. Steinhardt, Phys. Rev. Lett. \textbf{82}, 896 (1999).
\bibitem{ref:Chimento2003} L.P. Chimento, A. S. Jakubi, D. Pavon, and W. Zimdahl, Phys. Rev. D \textbf{67}, 083513 (2003). [arXiv:astro-ph/0303145]
\bibitem{ref:Huey2006} G. Huey and B.D. Wandelt, Phys. Rev. D \textbf{74}, 023519 (2006). [arXiv:astro-ph/0407196]

\bibitem{ref:Peebles2010} P.J.E. Peebles, AIP Conf. Proc. \textbf{1241}, 175 (2010).





\bibitem{ref:Marulli2012} F. Marulli, M. Baldi, and L. Moscardini, Mon. Not. Roy. Astron. Soc. \textbf{420}, 2377 (2012). [arXiv:1110.3045]
\bibitem{ref:Baldi2012} M. Baldi and P. Salucci, J. Cosmol. Astropart. Phys. \textbf{02} (2012) 014. [arXiv:1111.3953]
\bibitem{ref:Baldi2011} M. Baldi, Mon. Not. R. Astron. Soc. \textbf{414}, 116 (2011). [arXiv:1012.0002]
\bibitem{ref:Baldi2010} M. Baldi and M. Viel, Mon. Not. R. Astron. Soc. \textbf{409}, L89 (2010). [arXiv:1007.3736]
\bibitem{ref:Beynon2012} E. Beynon, M. Baldi, D. J. Bacon, K. Koyama, and C. Sabiu, Mon. Not. R. Astron. Soc. \textbf{422}, 3546 (2012). [arXiv:1111.6974]
\bibitem{ref:Lee2012} J. Lee and M. Baldi, Astrophys. J. \textbf{747}, 45 (2012). [arXiv:1110.0015]
\bibitem{ref:Cui2012} W. Cui, M. Baldi, and S. Borgani, Mon. Not. R. Astron. Soc. \textbf{424}, 993 (2012). [arXiv:1201.3568]

\bibitem{ref:Xia2013} J.-Q. Xia, J. Cosmol. Astropart. Phys. \textbf{11} (2013) 022. [arXiv:1311.2131]
\bibitem{ref:Xia2009} J.-Q. Xia, Phys. Rev. D \textbf{80}, 103514 (2009). [arXiv:0911.4820]
\bibitem{ref:Pourtsidou2013} A. Pourtsidou, C. Skordis, and E.J. Copeland, Phys. Rev. D \textbf{88}, 083505 (2013). [arXiv:1307.0458]
\bibitem{ref:Tarrant2012} E.R.M. Tarrant, C. van de Bruck, E.J. Copeland, and A.M. Green, Phys. Rev. D \textbf{85}, 023503 (2012). [arXiv:1103.0694]
\bibitem{ref:Ziaeepour2012} H. Ziaeepour, Phys. Rev. D \textbf{86}, 043503 (2012). [arXiv:1112.6025]
\bibitem{ref:Beyer2011} J. Beyer, S. Nurmi, and C. Wetterich, Phys. Rev. D \textbf{84}, 023010 (2011). [arXiv:1012.1175]
\bibitem{ref:Souza2010} R.C. de Souza and G.M. Kremer, Class. Quant. Grav. \textbf{27}, 175006 (2010). [arXiv:1006.3146]
\bibitem{ref:Vacca2009} G. La Vacca, J.R. Kristiansen, L.P.L. Colombo, R. Mainini, and S.A. Bonometto, J. Cosmol. Astropart. Phys. \textbf{04} (2009) 007. [arXiv:0902.2711]
\bibitem{ref:Manera2006} M. Manera and D.F. Mota, Mon. Not. R. Astron. Soc. \textbf{371}, 1373 (2006). [arXiv:astro-ph/0504519]
\bibitem{ref:Corasaniti2008} P.S. Corasaniti, Phys. Rev. D \textbf{78}, 083538 (2008). [arXiv:0808.1646]
\bibitem{ref:Mota2008} D.F. Mota, J. Cosmol. Astropart. Phys. \textbf{09} (2008) 006. [arXiv:0812.4493]

\bibitem{ref:Amendola2006} L. Amendola, M. Quartin, S. Tsujikawa, I. Waga, Phys. Rev. D \textbf{74}, 023525 (2006). [arXiv:astro-ph/0605488]
\bibitem{ref:Amendola2004} L. Amendola, Phys. Rev. D \textbf{69}, 103524 (2004). [arXiv:astro-ph/0311175]
\bibitem{ref:Amendola2000} L. Amendola, Phys. Rev. D \textbf{62}, 043511 (2000). [arXiv:astro-ph/9908023]
\bibitem{ref:Amendola2000-2} L. Amendola, Mon. Not. R. Astron. Soc. \textbf{312}, 521 (2000). [arXiv:astro-ph/9906073]
\bibitem{ref:Valentini2002} D. Tocchini-Valentini and L. Amendola, Phys. Rev. D \textbf{65}, 063508 (2002). [arXiv:astro-ph/0108143]
\bibitem{ref:Amendola1999} L. Amendola, Phys. Rev. D \textbf{60}, 043501 (1999). [arXiv:astro-ph/9904120]
\bibitem{ref:Wetterich1995} C. Wetterich, Astron. Astrophys. \textbf{301}, 321 (1995). [arXiv:hep-th/9408025]
\bibitem{ref:Holden2000} D.J. Holden and D. Wands, Phys. Rev. D \textbf{61}, 043506 (2000). [arXiv:gr-qc/9908026]
\bibitem{ref:Das2006} S. Das, P.S. Corasaniti, and J. Khoury, Phys. Rev. D \textbf{73}, 083509 (2006). [arXiv:astro-ph/0510628]
\bibitem{ref:Hwang2001} J.-C. Hwang and H. Noh, Phys. Rev. D \textbf{64}, 103509 (2001). [arXiv:astro-ph/0108197]
\bibitem{ref:Hwang2002} J.-C. Hwang and H. Noh, Class. Quant. Grav. \textbf{19}, 527 (2002). [arXiv:astro-ph/0103244]




\bibitem{ref:Potter2011} W.J. Potter and S. Chongchitnan, J. Cosmol. Astropart. Phys. \textbf{09} (2011) 005. [arXiv:1108.4414]
\bibitem{ref:Aviles2011} A. Aviles and J.L. Cervantes-Cota, Phys. Rev. D \textbf{84}, 083515 (2011). [arXiv:1108.2457]

\bibitem{ref:Caldera-Cabral2009} G. Caldera-Cabral, R. Maartens, and B.M. Schaefer, J. Cosmol. Astropart. Phys. \textbf{07} (2009) 027. [arXiv:0905.0492]
\bibitem{ref:Caldera-Cabral2009-2} G. Caldera-Cabral, R. Maartens, and L. A. Urena-Lopez, Phys. Rev. D \textbf{79}, 063518 (2009). [arXiv:0812.1827]
\bibitem{ref:Boehmer2010} C.G. Bohmer, G. Caldera-Cabral, N. Chan, R. Lazkoz, and R. Maartens, Phys. Rev. D \textbf{81}, 083003 (2010). [arXiv:0911.3089]
\bibitem{ref:Boehmer2008} C.G. Bohmer, G. Caldera-Cabral, R. Lazkoz, and R. Maartens, Phys. Rev. D \textbf{78}, 023505 (2008). [arXiv:0801.1565]
\bibitem{ref:Song2009} Y.-S. Song, L. Hollenstein, G. Caldera-Cabral, and K. Koyama, J. Cosmol. Astropart. Phys. \textbf{04} (2010) 018. [arXiv:1001.0969]
\bibitem{ref:Koyama2009} K. Koyama, R. Maartens, and Y.-S. Song, J. Cosmol. Astropart. Phys. \textbf{10} (2009) 017. [arXiv:0907.2126]

\bibitem{ref:Majerotto2010} E. Majerotto, J. Valiviita, and R. Maartens, Mon. Not. Roy. Astron. Soc. \textbf{402}, 2344 (2010). [arXiv:0907.4981]
\bibitem{ref:Valiviita2010} J. Valiviita, R. Maartens, and E. Majerotto, Mon. Not. Roy. Astron. Soc. \textbf{402}, 2355 (2010). [arXiv:0907.4987]

\bibitem{ref:Valiviita2008} J. Valiviita, E. Majerotto, and R. Maartens, J. Cosmol. Astropart. Phys. \textbf{07} (2008) 020. [arXiv:0804.0232]
\bibitem{ref:Jackson2009} B.M. Jackson, A. Taylor, and A. Berera, Phys. Rev. D \textbf{79}, 043526 (2009). [arXiv:0901.3272]
\bibitem{ref:Clemson2012} T. Clemson, K. Koyama, G.-B. Zhao, R. Maartens, and J. Valiviita, Phys. Rev. D \textbf{85}, 043007 (2012). [arXiv:1109.6234]
\bibitem{ref:Bean2008} R. Bean, E.E. Flanagan, and M. Trodden, Phys. Rev. D \textbf{78}, 023009 (2008). [arXiv:0709.1128]
\bibitem{ref:Bean2008-2} R. Bean, E.E. Flanagan, and M. Trodden, New J. Phys. \textbf{10}, 033006 (2008). [arXiv:0709.1124]


\bibitem{ref:Chongchitnan2009} S. Chongchitnan, Phys. Rev. D \textbf{79}, 043522 (2009). [arXiv:0810.5411]
\bibitem{ref:Gavela2009} M.B. Gavela, D. Hernandez, L. Lopez Honorez, O. Mena, and S. Rigolin, J. Cosmol. Astropart. Phys. \textbf{07} (2009) 034. [arXiv:0901.1611]


\bibitem{ref:Gavela2010} M.B. Gavela, L. Lopez Honorez, O. Mena, and S. Rigolin, J. Cosmol. Astropart. Phys. \textbf{11} (2010) 044. [arXiv:1005.0295]

\bibitem{ref:Salvatelli2013} V. Salvatelli, A. Marchini, L. Lopez-Honorez, and O. Mena, Phys. Rev. D \textbf{88}, 023531 (2013). [arXiv:1304.7119]
\bibitem{ref:Quartin2008} M. Quartin, M.O. Calvao, S.E. Joras, R.R.R. Reis, and I. Waga, J. Cosmol. Astropart. Phys. \textbf{05} (2008) 007. [arXiv:0802.0546]
\bibitem{ref:Honorez2010} L. Lopez Honorez, B.A. Reid, O. Mena, L. Verde, and R. Jimenez, J. Cosmol. Astropart. Phys. \textbf{09} (2010) 029. [arXiv:1006.0877]

\bibitem{ref:Costa2013} A.A. Costa, X.-D. Xu, B. Wang, E.G. Ferreira, and E. Abdalla, [arXiv:1311.7380].
\bibitem{ref:Bernardis2011} F. De Bernardis, M. Martinelli, A. Melchiorri, O. Mena, and A. Cooray, Phys. Rev. D \textbf{84}, 023504 (2011). [arXiv:1104.0652]
\bibitem{ref:He2011} J.-H. He, B. Wang, and E. Abdalla, Phys. Rev. D \textbf{83}, 063515 (2011). [arXiv:1012.3904]
\bibitem{ref:He2010} J.-H. He, B. Wang, E. Abdalla, and D. Pavon, J. Cosmol. Astropart. Phys. \textbf{12}, 022 (2010). [arXiv:1001.0079]
\bibitem{ref:He2008} J.-H. He and B. Wang, J. Cosmol. Astropart. Phys. \textbf{06} (2008) 010. [arXiv:0801.4233]
\bibitem{ref:Abdalla2009} E. Abdalla, L.R. Abramo, L. Sodre, and B. Wang, Phys. Lett. B \textbf{673}, 107 (2009). [arXiv:0910.5236]

\bibitem{ref:Sadjadi2010} H.M. Sadjadi, Eur. Phys. J. C \textbf{66}, 445 (2010). [arXiv:0904.1349]
\bibitem{ref:Olivares2008} G. Olivares, F. Atrio-Barandela, and D. Pavon, Phys. Rev. D \textbf{77}, 063513 (2008). [arXiv:0706.3860]
\bibitem{ref:Olivares2006} G. Olivares, F. Atrio-Barandela, and D. Pavon, Phys. Rev. D \textbf{74}, 043521 (2006). [arXiv:astro-ph/0607604]
\bibitem{ref:Olivares2005} G. Olivares, F. Atrio-Barandela, and D. Pavon, Phys. Rev. D \textbf{71}, 063523 (2005). [arXiv:astro-ph/0503242]

\bibitem{ref:Sun2013} C.-Y. Sun and R.-H. Yue, J. Cosmol. Astropart. Phys. \textbf{08} (2013) 018. [arXiv:1303.0684]
\bibitem{ref:Sadjadi2006} H.M. Sadjadi and M. Alimohammadi, Phys. Rev. D \textbf{74}, 103007 (2006). [arXiv:gr-qc/0610080]
\bibitem{ref:Sadeghi2013} J. Sadeghi, M. Khurshudyan, A. Movsisyan, and H. Farahani, J. Cosmol. Astropart. Phys. \textbf{12} (2013) 031. [arXiv:1308.3450]
\bibitem{ref:Zhang2013} M.-J. Zhang and W.-B. Liu, [arXiv:1312.0224].
\bibitem{ref:Koivisto2005} T. Koivisto, Phys. Rev. D \textbf{72}, 043516 (2005). [arXiv:astro-ph/0504571]
\bibitem{ref:Simpson2011} F. Simpson, B.M. Jackson, and J.A. Peacock, Mon. Not. R. Astron. Soc. \textbf{411}, 1053 (2011). [arXiv:1004.1920]
\bibitem{ref:Bertolami2007} O. Bertolami, F. Gil Pedro, and M. Le Delliou, Phys. Lett. B \textbf{654}, 165 (2007). [arXiv:astro-ph/0703462]
\bibitem{ref:Avelino2012} P.P. Avelino and H.M.R. da Silva, Phys. Lett. B \textbf{714}, 6 (2012). [arXiv:1201.0550]

\bibitem{ref:Quercellini2008} C. Quercellini, M. Bruni, A. Balbi, and D. Pietrobon, Phys. Rev. D \textbf{78}, 063527 (2008). [arXiv:0803.1976]
\bibitem{ref:Mohammadi2012} A. Khodam-Mohammadi and M. Malekjani, Gen. Rel. Grav. \textbf{44}, 1163 (2012). [arXiv:1101.1632]
\bibitem{ref:Sharif2012} M. Sharif and A. Jawad, Eur. Phys. J. C \textbf{72}, 2097 (2012). [arXiv:1212.0129]
\bibitem{ref:Fu2012} T.-F. Fu, J.-F. Zhang, J.-Q. Chen, and X. Zhang, Eur. Phys. J. C \textbf{72}, 1932 (2012). [arXiv:1112.2350]
\bibitem{ref:Li2011} Y.-H. Li and X. Zhang, Eur. Phys. J. C \textbf{71}, 1700 (2011). [arXiv:1103.3185]
\bibitem{ref:Barrow2006} J.D. Barrow and T. Clifton, Phys. Rev. D \textbf{73}, 103520 (2006). [arXiv:gr-qc/0604063]
\bibitem{ref:Zimdahl2001} W. Zimdahl and D. Pavon, Phys. Lett. B \textbf{521}, 133 (2001). [arXiv:astro-ph/0105479]


\bibitem{ref:Lip2011d} S.Z.W. Lip, Phys. Rev. D \textbf{83}, 023528 (2011). [arXiv:1009.4942]
\bibitem{ref:Chen2011} X. Chen, B. Wang, N. Pan, and Y. Gong, Phys. Lett. B \textbf{695}, 30 (2011). [arXiv:1008.3455]
\bibitem{ref:Chen2009} X. Chen, Y. Gong, and E.N. Saridakis, J. Cosmol. Astropart. Phys. \textbf{04} (2009) 001. [arXiv:0812.1117]
\bibitem{ref:Koshelev2009} N.A. Koshelev, Gen. Rel. Grav. \textbf{43}, 1309 (2011). [arXiv:0912.0120]
\bibitem{ref:Zhang2012} Z. Zhang, S. Li, X.-D. Li, X. Zhang, and M. Li, J. Cosmol. Astropart. Phys. \textbf{06} (2012) 009. [arXiv:1204.6135]
\bibitem{ref:Cao2011} S. Cao, N. Liang, and Z.-H. Zhu, Mon. Not. Roy. Astron. Soc. \textbf{416}, 1099 (2011). [arXiv:1012.4879]
\bibitem{ref:Guo2007} Z.-K. Guo, N. Ohta, and S. Tsujikawa, Phys. Rev. D \textbf{76}, 023508 (2007). [arXiv:astro-ph/0702015]


\bibitem{ref:Liyh2013} Y.-H. Li and X. Zhang, [arXiv:1312.6328]
\bibitem{ref:Bolotin2013} Y.L. Bolotin, A. Kostenko, O.A. Lemets, and D.A. Yerokhin, [arXiv:1310.0085].
\bibitem{ref:Chimento2013} L.P. Chimento, M.G. Richarte, and Ivan E. Sanchez Garcia, Phys. Rev. D \textbf{88}, 087301 (2013). [arXiv:1310.5335]
\bibitem{ref:Chimento2012RDE} L.P. Chimento and M.G. Richarte, Phys. Rev. D \textbf{85}, 127301 (2012). [arXiv:1207.1492]
\bibitem{ref:Chimento2011RDE} L.P. Chimento and M.G. Richarte, Phys. Rev. D \textbf{84}, 123507 (2011). [arXiv:1107.4816]
\bibitem{ref:Tong2011} M.-L. Tong, Y. Zhang, and Z.-W. Fu, Class. Quant. Grav. \textbf{28}, 055006 (2011). [arXiv:1101.5199]



\bibitem{ref:fsigma81-Percival2004} W.J. Percival \textit{et al.} [The 2dFGRS Collaboration], Mon. Not. Roy. Astron. Soc. \textbf{353}, 1201 (2004).
\bibitem{ref:fsigma82-Blake2011} C. Blake \textit{et al.}, Mon. Not. Roy. Astron. Soc. \textbf{415}, 2876 (2011).
\bibitem{ref:fsigma83-Samushia2012} L. Samushia, W.J. Percival, and A. Raccanelli, Mon. Not. Roy. Astron. Soc. \textbf{420}, 2102 (2012).
\bibitem{ref:fsigma84-Reid2012} B.A. Reid \textit{et al.}, Mon. Not. Roy. Astron. Soc. \textbf{426}, 2719 (2012). [arXiv:1203.6641]
\bibitem{ref:fsigma85-Beutler2012} F. Beutler \textit{et al.}, Mon. Not. Roy. Astron. Soc. \textbf{423}, 3430 (2012). [arXiv:1204.4725]
\bibitem{ref:fsigma8total-Samushia2013} L. Samushia, \textit{et al.}, Mon. Not. Roy. Astron. Soc. \textbf{429}, 1514 (2013). [arXiv:1206.5309]
\bibitem{ref:fsigma86-Torre2013} S. de la Torre, \textit{et al.}, Astron. Astrophys. \textbf{557}, A54 (2013). [arXiv:1303.2622]
\bibitem{ref:fsigma87-Macaulay2013} E. Macaulay, I.K. Wehus, and H.K. Eriksen, Phys. Rev. Lett. \textbf{111}, 161301 (2013).

\bibitem{ref:fsigma8-DE-Song2009} Y.-S. Song and W.J. Percival, J. Cosmol. Astropart. Phys. \textbf{10} (2009) 004. [arXiv:0807.0810]
\bibitem{ref:fsigma8-HDE-Xu2013} L. Xu, Phys. Rev. D \textbf{87}, 043525 (2013). [arXiv:1302.2291]
\bibitem{ref:fsigma8andPlanck-MG-Xu2013} L. Xu, Phys. Rev. D \textbf{88}, 084032 (2013). [arXiv:1306.2683]
\bibitem{ref:Xu2013-DGP} L. Xu, J. Cosmol. Astropart. Phys. \textbf{02} (2014) 048. [arXiv:1312.4679].
\bibitem{ref:Yang2013-1} W. Yang and L. Xu, [arXiv:1311.3419].
\bibitem{ref:Yang2013-2} W. Yang, L. Xu, Y. Wang, and Y. Wu, Phys. Rev. D \textbf{89}, 043511 (2014). [arXiv:1312.2769]
\bibitem{ref:Yang2014-ux} W. Yang and L. Xu, [arXiv:1401.5177].

\bibitem{ref:BAO-1} F. Beutler, \textit{et al.}, Mon. Not. Roy. Astron. Soc. \textbf{416}, 3017 (2011). [arXiv:1106.3366]
\bibitem{ref:BAO-2} N. Padmanabhan, \textit{et al.}, Mon. Not. Roy. Astron. Soc. \textbf{427}, 2132 (2012). [arXiv:1202.0090]
\bibitem{ref:BAO-3} M. Manera, \textit{et al.}, Mon. Not. Roy. Astron. Soc. \textbf{428}, 1036 (2013). [arXiv:1203.6609]

\bibitem{ref:SNLS3-1} J. Guy, \textit{et al.}, Astron. Astrophys. \textbf{523}, A7 (2010). [arXiv:1010.4743]
\bibitem{ref:SNLS3-2} A. Conley, \textit{et al.}, Astrophys. J. Suppl. \textbf{192}, 1 (2011).
\bibitem{ref:SNLS3-3} M. Sullivan, \textit{et al.}, Astrophys. J. \textbf{737}, 102 (2011).


\bibitem{ref:Ma1995} C.P. Ma and E. Berschinger, Astrophys. J. \textbf{455}, 7 (1995).

\bibitem{ref:Kodama1984} H. Kodama and M. Sasaki, Prog. Theor. Phys. \textbf{78}, 1 (1984).
\bibitem{ref:Hu1998} W. Hu, Astrophys. J. \textbf{506}, 485 (1998). [arXiv:astro-ph/9801234]
\bibitem{ref:Gordon2004} C. Gordon and W. Hu, Phys. Rev. D \textbf{70}, 083003 (2004). [arXiv:astro-ph/0406496]

\bibitem{ref:Caldwell2003} R.R. Caldwell, M. Kamionkowski, and N.N. Weinberg, Phys. Rev. Lett. \textbf{91}, 071301 (2003).


\bibitem{ref:Linder2003} E.V. Linder and A. Jenkins, Mon. Not. Roy. Astron. Soc. \textbf{346}, 573 (2003).

\bibitem{ref:Tsujikawa2010} S. Tsujikawa, Lect. Notes Phys. \textbf{800}, 99 (2010).


\bibitem{ref:camb} A. Lewis, A. Challinor, and A. Lasenby, Astrophys. J. \textbf{538}, 473 (2000); http://camb.info/.
\bibitem{ref:cosmomc-Lewis2002} A. Lewis and S. Bridle, Phys. Rev. D \textbf{66}, 103511 (2002); http://cosmologist.info/cosmomc/.


\bibitem{ref:Hu1995} W. Hu, PhD Thesis, University of California at Berkeley, arXiv:astro-ph/9508126.


\bibitem{ref:WMAP7} E. Komatsu \textit{et al.}, Astrophys. J. Suppl. Ser. \textbf{192}, 18 (2011).
\bibitem{ref:BAO-Clemson2012} W.J. Percival \textit{et al.}, Mon. Not. R. Astron. Soc. \textbf{381}, 1053 (2007).
\bibitem{ref:HST-Clemson2012} A.G. Riess, Astrophys. J. \textbf{699}, 539 (2009).
\bibitem{ref:SNIa-Clenmson2012} R. Kessler \textit{et al.}, Astrophys. J. Suppl. Ser. 185, 32 (2009).

\bibitem{ref:HST-Salvatelli2013} A.G. Riess \textit{et al.}, Astrophys. J. \textbf{730}, 119 (2011); \textbf{732}, 129(E) (2011).

\bibitem{ref:Eisenstein2005} D.J. Eisenstein \textit{et al.}, Astrophys. J. 633, 560 (2005).
\bibitem{ref:Percival2007} W.J. Percival, S. Cole, D.J. Eisenstein, R.C. Nichol, J.A. Peacock, A.C. Pope, and A.S. Szalay, Mon. Not. R. Astron. Soc. \textbf{381}, 1053 (2007).

\bibitem{ref:Percival2009} W.J. Percival, and M. White, Mon. Not. R. Astron. Soc. \textbf{393}, 297 (2009).



\end{thebibliography}
\end{document}